\documentclass[a4paper,fleqn,usenatbib]{mnras}

\usepackage{graphicx}
\usepackage{amssymb}
\usepackage{amsmath}
\usepackage{times} 
\usepackage{multirow}
\usepackage{url}

\usepackage{hyperref}
\usepackage{array}

\title[Growth rate in the local Universe with ALFALFA]{The growth rate of cosmic structures in the local Universe with the ALFALFA survey}

\author[Avila et al.]{F. Avila,$ ^{1} $\thanks{e-mail: felipeavila@on.br}
A. Bernui,$^{1}$
E. de Carvalho,~$\!^{2}$ C. P. Novaes$^{3}$\\
$^{1}$Observat\'orio Nacional, Rua General Jos\'e Cristino 77, 
S\~ao Crist\'ov\~ao, 20921-400 Rio de Janeiro, RJ, Brazil \\
$^{2}$Centro de Estudos Superiores de Tabatinga, Universidade do Estado do Amazonas, 
69640-000, Tabatinga, AM, Brazil \\
$ ^{3} $Instituto Nacional de Pesquisas Espaciais, Av. dos Astronautas 1758, Jardim da Granja, S\~ao Jos\'e dos Campos, SP, Brazil
}

\date{Accepted XXX. Received YYY; in original form ZZZ}

\pubyear{2021}

\begin{document}

\maketitle

\begin{abstract}
We investigate the growth rate of structures in the local Universe. 
For this, we use as a cosmological tracer the HI line extra-galactic sources from the Arecibo 
Legacy Fast ALFA (ALFALFA) survey to obtain a measurement of the normalized growth 
rate parameter, $f \sigma_{8}$, considered a powerful tool to constrain alternative models 
of gravity. 
For these analyses, we calculate the Local Group velocity due to the matter structures 
distribution in the ALFALFA catalogue and compare it with the Local Group velocity 
relative to the Cosmic Microwave Background frame to obtain the velocity scale parameter, 
$\beta$. 
Using Monte Carlo realizations and log-normal simulations, our methodology quantifies the 
errors introduced by shot-noise and partial sky coverage of the analysed data. 
The measurement of the velocity scale parameter $\beta$, and the calculation of the 
matter fluctuation of the cosmological tracer, $\sigma_{8}^{\text{tr}}$, lead us to 
$f \sigma_{8} = 0.46 \pm 0.06$ at $\bar{z} = 0.013$, in good agreement (at $1 \sigma$ 
level) with the value expected in the $\Lambda$CDM concordance model. 
In addition, our analyses of the ALFALFA sample also provide a 
measurement of the growth rate of structures $f \,=\, 0.56 \pm 0.07$, at $\bar{z} = 0.013$.

\end{abstract}


\section{Introduction}\label{sec1}

The large-scale structure data from current astronomical surveys contain the imprints of 
matter clustering evolution caused by gravitational 
instability~\citep{Pezotta,BAODR14,Haude19,Marques20,Aubert20}. 
How these instabilities evolve over time is a crucial test for theories that aim to understand 
possible deviations of cosmological models based on general relativity (GR). 
The growth rate of cosmic structures, $f(a)$, is defined as \citep{Strauss}
\begin{equation}\label{growth rate}
f(a) \equiv \dfrac{d\ln D(a)}{d\ln a} \, ,
\end{equation}
where $D = D(a)$ is the linear growth function, and $a$ is the scale factor in the 
Robertson-Walker metric based on GR theory. 
In fact, $f = f(a)$ has the potential to constrain alternative models of gravity and dark 
energy from the measure of the growth index, $\gamma$, when one parametrizes $f$ 
as~\citep{Linder} 
\begin{equation}\label{growth rate parametrize}
f(a) = \Omega^{\gamma}_{m} \, .
\end{equation}
In the $\Lambda$CDM model, based on GR, $\gamma \simeq 0.55$, and for some $f(R)$ 
modified gravity models $\gamma \simeq 0.41 - 0.21z$~\citep{Basilakos12}.

The literature reports the analyses of several cosmological probes used to constrain $f(a)$, 
like the Cosmic Microwave Background (CMB)~\citep{Planck18}, cluster 
abundances~\citep{Planck16}, weak lensing~\citep{DES}, redshift space distortions 
(RSD)~\citep{Alam17}, and peculiar velocities~\citep{Boruah}, among others.
Measurements of the growth rate, as well as other cosmological observables, are useful to determine the best parameters of the current cosmological model \citep{deCarvalho18,Nunes20}. 
 
Although some studies intend to measure $f(a)$ at some redshift, in practice they also 
constrain other parameters. 
For instance, the approach with peculiar velocities constrains the velocity scale parameter 
$\beta \equiv f/b$, where $b$ is the linear bias defined by 
$b \equiv \sigma_{8}^{\text{tr}} / \sigma_{8}^{\text{m}}$~\citep{Papageorgiou}, 
$\sigma_{8}^{m}$ is the matter fluctuation at radius 8 Mpc\,/$h$ (hereafter 
$\sigma_{8} \equiv \sigma_{8}^{m}$) and $\sigma_{8}^{\text{tr}}$ is the matter fluctuation 
of the cosmological tracer (e.g., blue galaxies, luminous red galaxies, etc.) 
at radius 8 Mpc\,/$h$. 
The most common constraint found in the literature is the combination 
$f \sigma_{8}$. 
An interesting approach to obtain $f\sigma_{8}$ is to measure first $\beta$, and then find 
$f\,\sigma_{8} = \beta\,b\,\sigma_{8}$, or equivalently 
\begin{equation}\label{fsig8}
f \, \sigma_{8} = \beta \, \sigma_{8}^{\text{tr}} \, .
\end{equation}

For a local Universe sample, $z \lesssim 0.1$, the gravitational dipole approach is suitable to 
measure $\beta$ and, consequently, $f\,\sigma_{8}$, using the equation (\ref{fsig8}) (see, 
e.g.,~\cite{Strauss,Scaramella}). 
If one compares the peculiar velocity of our Local Group (LG) of galaxies inferred from the 
CMB dipole, $627 \pm 22$ km\,/\,s towards 
$(l, b) = (273^{\circ} \pm 3^{\circ}, 29^{\circ}\pm3^{\circ})$~\citep{Kogut,Courteau,Erdogdu}, 
with the value measured in a local survey, one can constrain $\beta$. 
For example, in the analyses done by~\cite{Erdogdu}, they used the 2 Micron All-sky Redshift 
Survey (2MRS) to estimate $\beta = 0.40 \pm 0.09$, with the dipole converging to a constant 
value around 60 Mpc\,/$h$. 
In another work,~\cite{Basilakos06} re-examined the Point Source catalogue redshift 
(PSCz) survey \citep{rowan} estimating the $\beta$ parameter as $\beta \simeq 0.49$, with significantly 
contribution to the dipole magnitude from distances beyond 185~Mpc\,/$h$. 
In addition, using a full-sky X-ray cluster sample, \cite{Kocevski} 
calculated $\beta = 0.24\pm0.01$ using the number-weighted method (see Section~\ref{sec3.1}). 
These authors also observe contributions to the dipole velocity beyond 185 Mpc\,/$h$. As reported by \cite{Bilicki11}, one should not expect consistency between different analyses of diverse cosmic tracers for the amplitude and the scale of convergence. Regarding $\beta$, the $\Lambda$CDM model is a good guide to find out if the result found for $\beta$ is in any way consistent, that is, $\beta = \Omega_m^{\Lambda\text{CDM}}(z)^{0.55}/ b$. Despite the differences found in the literature about the convergence scale, that is, the 
minimum scale where the dipole velocity attains its stability 
value~\citep{Kocevski}, 
we will show, for the sample in analysis, that the magnitude of the gravitational dipole is reached around 60~Mpc. 

In this work we investigate the clustering of the extra-galactic HI line sources observed in the, 
recently completed, ALFALFA Survey~\citep{Haynes18} to perform a measurement of 
$f \sigma_{8}$ in the local Universe, i.e., at $\bar{z}=0.013$. 
Although the ALFALFA catalogue does not contain full-sky data, since the surveyed area is 
$\Omega \simeq 6900 ~\text{deg}^{2}$ ($f_{sky} \simeq 1/6$), a distinctive feature is that it 
has a high number density of objects compared with full-sky catalogues in the same redshift 
range. 
This reduces the shot noise and increases the efficiency of the selection function, ruling out 
artificial convergences in the dipole magnitude, a crucial 
property for the approach we adopt here, and that we shall explain below. 
Another important attribute of the ALFALFA catalogue is that for the data sample 
with $c z_{\odot} < 6000$ km\,/\,s the dipole can be calculated 
in the real space avoiding the RSD effect, known as \textit{the rocket effect} 
\citep{Kaiser87,Kaiser89}. 
Regarding the error and systematic sources, our analyses take into account the incomplete 
sky coverage by using log-normal simulations~\citep{Agrawal17}, that help us to correct 
both direction and magnitude of the gravitational dipole 
(also called clustering dipole), and a set of Monte Carlo realizations, proposed 
by \cite{Basilakos98} to estimate the shot-noise error, procedures described in 
Section~\ref{sec3.2} 
\citep[for diverse analyses regarding systematics see, e.g.,][]{Marques18,Avila18,Avila19,%
deCarvalho20,deCarvalho21,Sarkar19, Pandey20, Heinesen20}. 

The work is structured as follows. 
The ALFALFA catalogue is presented in Section~\ref{sec2}, together with the selection 
function calculation and the criteria to select the final data sample for analysis. 
In Section~\ref{sec3} we detail the methodology to calculate the LG dipole, their error 
estimation and the $\sigma_{8}^{\text{tr}}$ calculation. 
The results of our analyses and our conclusions are presented in Sections~\ref{sec4} 
and~\ref{sec5}, respectively.

\section{The Arecibo Legacy Fast ALFA Survey}\label{sec2}

The Arecibo Legacy Fast ALFA 
Survey\footnote{\url{http://egg.astro.cornell.edu/alfalfa/data/index.php}} 
\citep[ALFALFA;][]{Haynes11,Haynes18} was a blind 21 cm HI line survey designed with 
the main goal of obtaining a robust measurement of the HI mass function, an important 
component, together with the luminosity function, that can make a significant contribution to 
the study of galaxy population in the local 
Universe~\citep{Giovanelli,Jones16,Jones18,Jones20}. 
Given the surveyed area and the spectral resolution, ALFALFA can measure the faint end 
of the HI mass function for the optically faint, gas-rich population \citep{Donoghue}. 
For additional studies see~\cite{Haynes18} and references therein.

The ALFALFA survey was performed between 2005 and 2011 covering an area of 
$\Omega\simeq 6 900~\text{deg}^{2}$ out to $z < 0.06$ detecting $31 500$ HI line 
extra-galactic sources. 
The survey covers two discontinuous regions, both in the declination range of 
$0^{\circ} < \text{DEC} < 36^{\circ}$, in the right ascension intervals of 
$21^{\text{h}} 30^{\text{m}} < \text{RA} < 3^{\text{h}} 15^{\text{m}}$ 
(South galactic hemisphere) 
and $7^{\text{h}} 20^{\text{m}} < \text{RA} < 16^{\text{h}}40^{\text{m}}$ 
(North galactic hemisphere). 
The catalogue distinguish the sources with a {\sc code} 1, 2, and 9, according to quality of 
the data observed~\citep{Haynes18}. 
{\sc code} 1, refers to a high signal-to-noise ratio detection of the HI extragalactic source, 
with confirmed optical counterpart; 
{\sc code} 2, lower signal to noise ratio coincident with optical counterpart, they are 
considered unreliable sources; and 
{\sc code} 9, high signal to noise ratio source with no optical counterpart and likely Galactic 
high velocity cloud. 
In this work we shall use only the sources with {\sc code 1}, as recommended by the 
ALFALFA team.

\subsection{Data selection}\label{sec2.1}

The distances data presented in the ALFALFA catalogue are described 
in~\cite{Haynes18} (see Section 3.1, column 11). 
The ALFALFA collaboration uses two distances estimation approaches: 
(i) for those objects with $cz_{\odot} > 6000$ km/s the distance is simply estimated as 
$c\,z_{\text{CMB}}/H_0$, 
where $c\,z_{cmb}$ is the recessional velocity measured in the Cosmic Microwave 
Background reference frame and $H_0$ is the Hubble constant; and, 
(ii) for objects with $cz_{\text{CMB}} < 6000$ km/s the collaboration assigns distances to 
nearby galaxies through a parametric flow model developed by~\cite{Master}, 
based mainly on the SFI++ catalogue of galaxies~\citep{Spring07} and results from 
analysis of the peculiar motion of galaxies, groups, and clusters, using a combination of 
primary distances from the literature and secondary distances from the Tully-Fisher relation. 
Also, when available, they use known distances from the literature. In Appendix~\ref{apendiceA} we test the impact of distance uncertainties on dipole analyses.

\begin{figure}
\centering
\includegraphics[width=8.0cm,height=5.0cm]{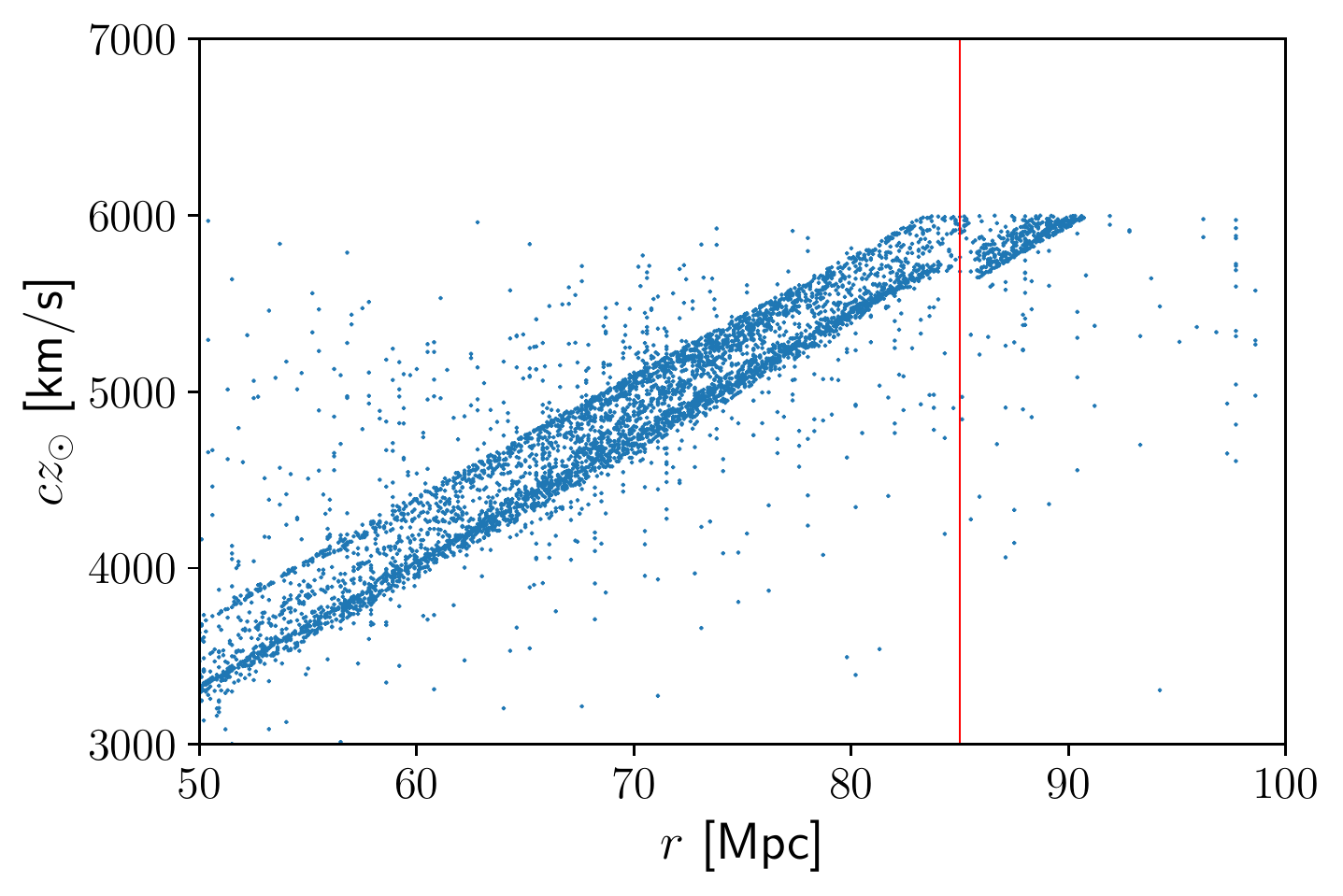}
\caption{The Hubble-Lema\^{\i}tre diagram for the ALFALFA HI line sources below $cz_{\odot} < 6000$ km/s 
and with {\sc code} 1. Notice the discontinuity around $r = 85$ Mpc, which appears because the ALFALFA team adopted a different approach to determine the distances. 
For this, we decided to remove from our sample those objects at distances larger than 85 
Mpc.}
\label{fig:cut85}
\end{figure}

The transition velocity between the methodologies applied to calculate distances, that is, 
$6000$ km/s, means $85 - 90$ Mpc, and this interval corresponds to the discontinuity 
observed in the Hubble-Lema\^{\i}tre diagram, shown in Fig.~\ref{fig:cut85}. 
Due to this fact, we applied a conservative cut to remove the HI sources with distances 
above 85 Mpc (the red vertical line in Fig. \ref{fig:cut85} illustrates this cut).
In this way we are sure that all objects from the sample selected for our dipole analyses 
have all their distance measurements performed using only one methodology. As also observed in Fig.~\ref{fig:cut85}, some objects appear far from the Hubble flow, probably due to their intense peculiar motions. This systematic effect motivates an examination. For this, in Appendix~\ref{apendiceB}, we performed a comparative test with and without the outlier HI sources.

\begin{figure}
	\centering
	\includegraphics[width=8.0cm,height=5.0cm]{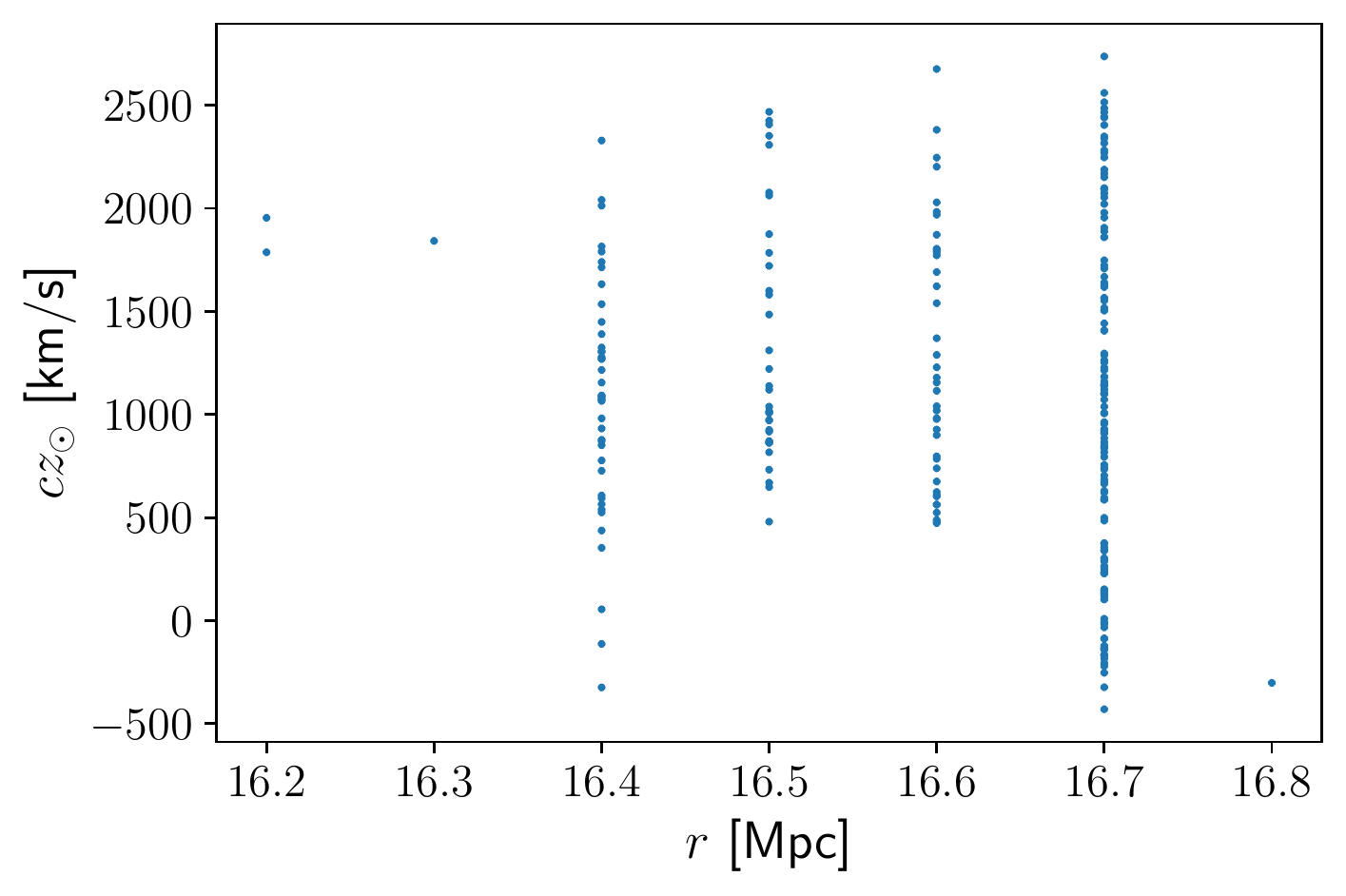}
\caption{Distribution of distances {\it versus} velocities (i.e., the Hubble-Lema\^{\i}tre plot) 
for the members of the Virgo cluster. 
One observes various discontinuities, evidencing the difficulty of the ALFALFA survey to 
determine the individual distances of the Virgo cluster galaxies, suggestive of non-linear 
dynamics that can bias the analyses. These objects were removed from the sample in 
analysis.}
\label{fig:virgovel}
\end{figure}

The next constraint applied to the sample concerns the presence of the Virgo cluster. 
We observe that it is composed by 224 HI line sources, a potential source of systematics. 
In fact, the radial distances attributed to the members of the Virgo cluster in the ALFALFA 
catalogue is not realistic, as observed in Fig. \ref{fig:virgovel} \citep[see discussions about 
the adopted distances in][]{Haynes18}. 
The discontinuous distribution of distances of the galaxy Virgo members, 
clearly observed in Fig. \ref{fig:virgovel}, is due to the lack of information regarding 
the distance of each galaxy to us. 
For this, the distances of the galaxies inside Virgo are only assigned by identifying the 
groups or substructures and assigning the same distance to all their members.
Therefore, to avoid biasing our calculations, we remove the Virgo members from 
our sample.

Our final restriction aims to avoid the contribution of LG member galaxies, which should 
not be accounted for the dipole calculation. 
Therefore, since we are considering the LG as an unique structure, we 
follow~\cite{Bilicki11,Bilickithesis,Erdogdu} and remove these galaxies. 
We assume the LG as a spherical structure of 1.5 Mpc radius centred near us, as 
suggested by observations~\citep{Berghbook,Marel}. 
Then, we localize and remove from our catalogue 8 HI line sources corresponding to this 
region.

\begin{figure}
\centering
\includegraphics[width=8.0cm,height=5.0cm]{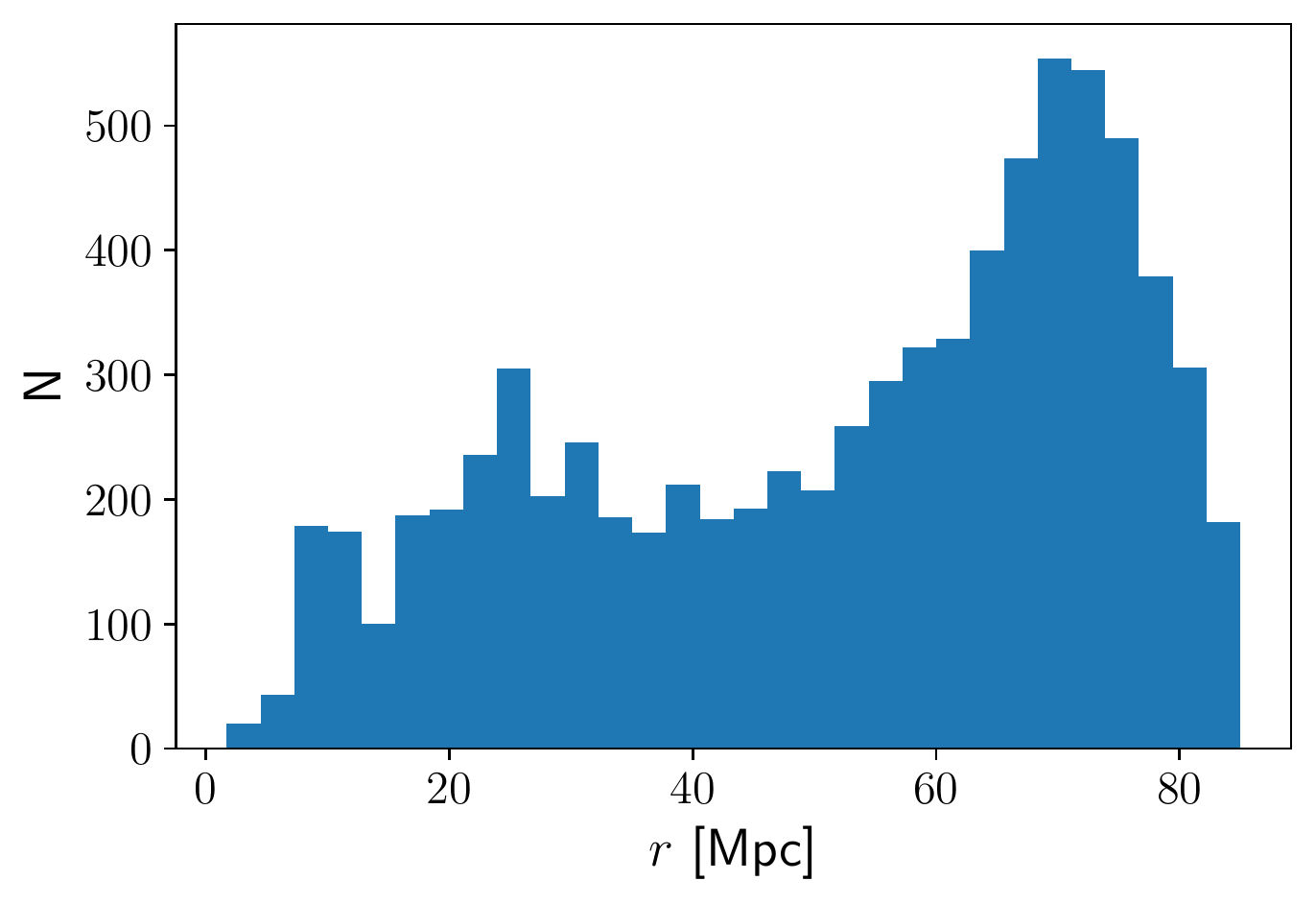}
\caption{Histogram of the distances distribution of the ALFALFA HI line sources of 
the final data sample used in this work. See Section \ref{sec2.1} for more information on how 
the sample was selected.}
\label{fig:hist}
\end{figure}

\begin{figure}
	\centering
	\includegraphics[width=8.0cm,height=5.0cm]{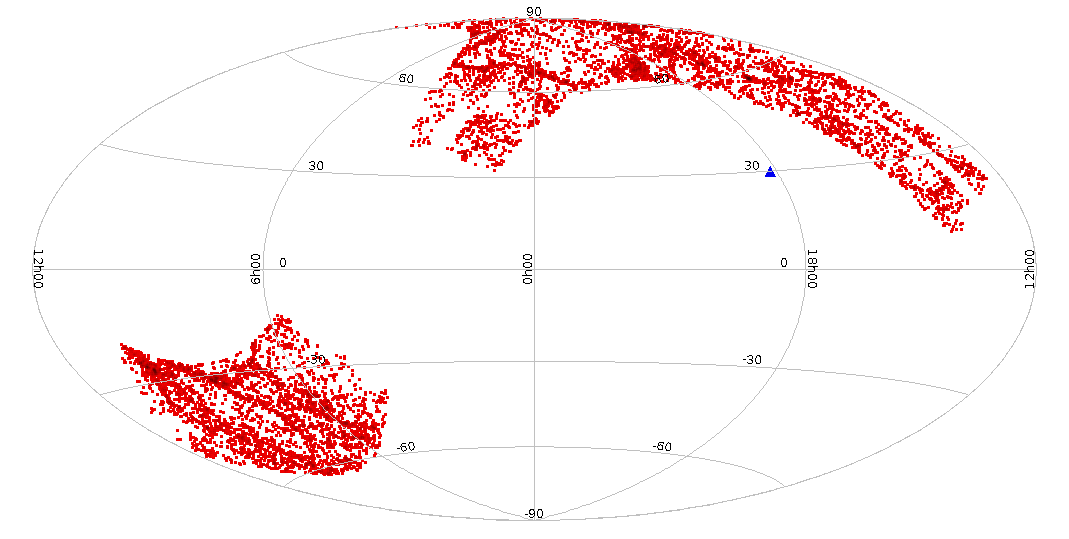}
	\caption{The HI line sources of our final data sample in the galactic Aitoff projection. 
The blue triangle indicates the direction of the LG velocity $(l, b) = (273^{\circ} \pm 3^{\circ}, 29^{\circ} \pm3^{\circ})$ inferred using the CMB dipole measurement.
}
\label{fig:skyplot}
\end{figure}

After all these cuts the final data sample for analyses contains $N = 7798$ HI line 
extra-galactic sources, with median redshift $\bar{z} = 0.013$, and number density 
$\bar{n} = 0.04$ Mpc$^{-3}$. 
In Fig. \ref{fig:hist} we show the distribution of their distances, and in Fig. \ref{fig:skyplot} 
their Aitoff projection on the celestial sphere in galactic coordinates.


Two important features of this final data sample regarding our study of the gravitational dipole 
deserve some comments: 
(i) it does not cover the full sky, and 
(ii) it does not cover the area corresponding to the LG peculiar motion direction inferred 
from the CMB dipole, as shown in Fig. \ref{fig:skyplot} as a blue triangle. 
With respect to the first feature, our methodology to estimate the error for $f \sigma_{8}$ takes into account the incomplete sky coverage, as described in the Section \ref{sec3.2}. 
Regarding the second feature, one does not expect that with a partial sky coverage of the data 
sample one could find the LG dipole direction aligned, or close to, the CMB dipole direction; 
fortunately, our interest is just in the modulus part of the LG dipole, 
which is affected by the partial coverage of the sky, but can be corrected by evaluating the 
impact of such feature through a set of simulated data, as we shall see.

\subsection{The radial selection function}\label{sec2.2}

\begin{figure}
\centering
\includegraphics[width=8.0cm,height=5.0cm]{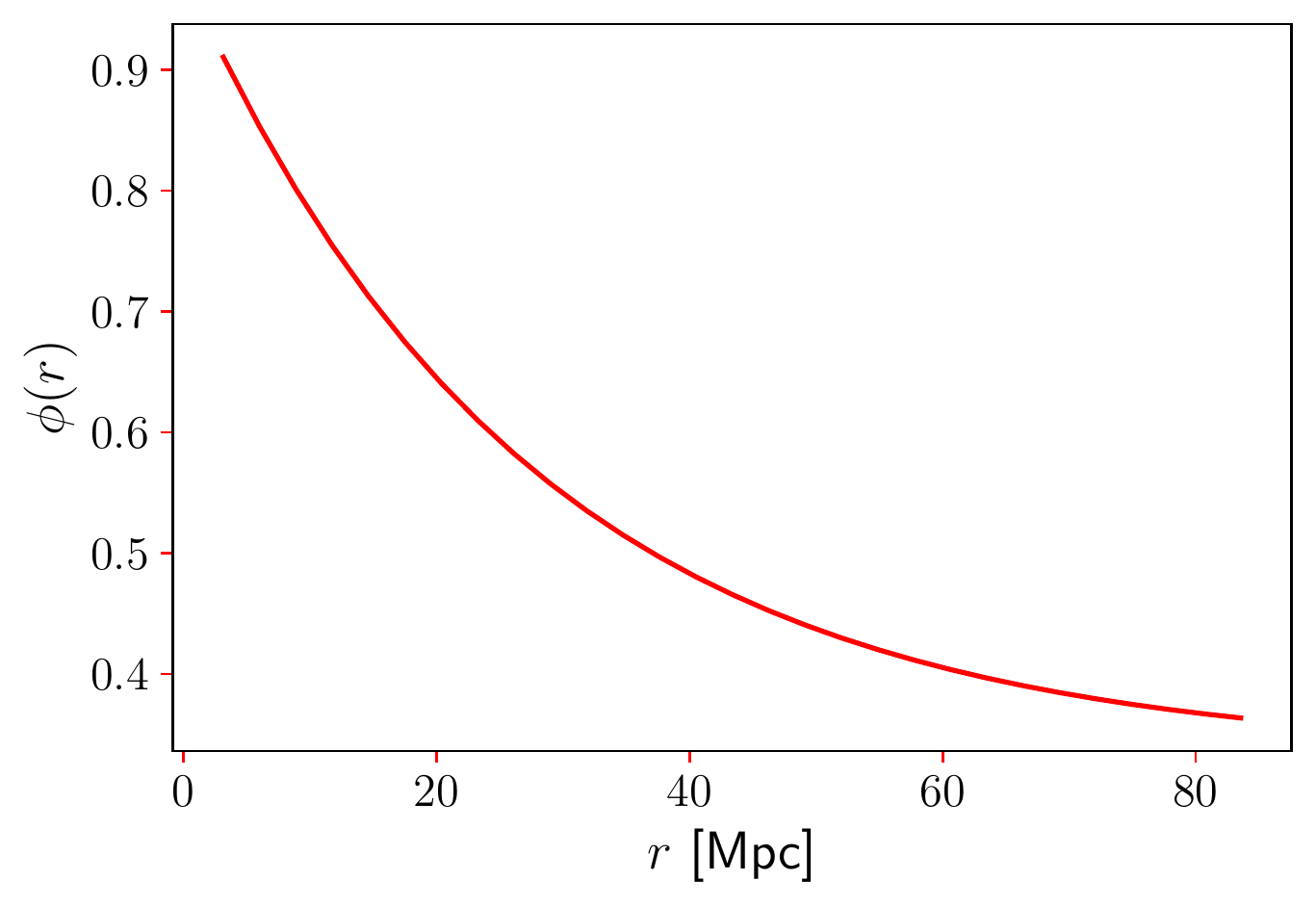}
\caption{Radial selection function for the data in analysis. For details, see 
Section~\ref{sec2.2}.}
\label{fig:selectionfunction}
\end{figure}

In order to calculate the gravitational dipole we must assign a weight $\omega_{i}$ 
for the $i\:\!$th galaxy~\citep{Scaramella}
\begin{equation}\label{weight}
\omega_{i} = \frac{1}{\phi(r_{i})} \, ,
\end{equation}
where $\phi(r_i)$ is the radial selection function. 
To calculate $\phi(r)$ for the ALFALFA sample we must obtain the two-dimensional 
number density distribution, $n(m_{\text{HI}}, \omega_{50})$, where $m_{\text{HI}} \equiv \log(M_{\text{HI}})$ and $M_{\text{HI}}$ is the HI mass, measured in solar mass units, and $w_{50} \equiv log(W_{50})$ is associated to the velocity width of the HI spectrum, $W_{50}$, measured at the 50\% level, as described in \cite{Haynes18}, in units of km/s. 
Then, $\phi(r)$ is calculated as~\citep{Papastergis}
\begin{equation}\label{selection_function}
\phi(r) = \dfrac{\int_{\omega_{i}}^{\omega_{f}}\int_{m(r)}^{m_{f}} n(m, \omega) dm\,d\omega}{\int_{\omega_{i}}^{\omega_{f}}\int_{m_{i}}^{m_{f}} n(m, \omega) dm\,d\omega}.
\end{equation}
We dropped the sub-indices HI and 50 for simplicity. 
Notice that the integral in mass in the numerator uses as lower limit the minimum HI mass detectable at a distance $r$, $m_{\text{HI}}(r)$. The other integrals are performed over the whole range of mass $m_{i} \leqslant m_{\text{HI}} \leqslant m_{f}$, and velocity width, $\omega_{i} \leqslant \omega_{50} 
\leqslant \omega_{f}$, of the objects in the ALFALFA sample.
In Fig. \ref{fig:selectionfunction} we show the result of this calculation, that is, the radial 
selection function for our data analyses. 

\section{Methodology}\label{sec3}

In this section we discuss \textit{the number-weighted method}, a procedure to measure 
the LG velocity, caused by large-scale matter distribution, and the $\beta$ parameter, an 
intermediary step to obtain $f\sigma_{8}$. 
To apply this methodology one assumes the data to be a representative sample of the 
local Universe, besides that the galaxy clustering can be analysed using linear theory of 
perturbations. 
In what follows, we also describe the error estimations, that is, the contribution from 
shot-noise and the incomplete sky coverage, and the necessary correction. 
Lastly, we show the way to measure $\sigma_{8}^{\text{tr}}$ to obtain $f\sigma_{8}$ from equation (\ref{fsig8}).

\subsection{Number-weighted method and the LG dipole}\label{sec3.1}

From the linear theory of gravitational instability, the peculiar velocity field \textbf{v} is 
related to the gravitational field, \textbf{g}, by \citep{peebles}
\begin{equation}\label{relationvelocitygravity}
\textbf{v}(\textbf{r}) = \dfrac{H_{0}f}{4\pi G\bar{\rho}} \, 
\textbf{g}(\textbf{r}) \, ,
\end{equation}
where $f$ is the growth rate defined in equation (\ref{growth rate}), $H_{0}$ is 
the Hubble constant measured today, $G$ is the gravitational constant, and $\bar{\rho}$ 
is the matter density averaged over a large volume $V$. 
If we write $\textbf{g}$ as
\begin{equation}\label{equationforg}
\textbf{g}(\textbf{r}) = 
G \bar{\rho} \int_{V}\delta_{\text{m}}(\textbf{r}')\dfrac{\textbf{r}'-\textbf{r}}{|\textbf{r}'-\textbf{r}|^{3}} \, d^{3}\textbf{r}' \, ,
\end{equation}
one can calculate the peculiar velocity if one knows the matter density contrast, 
$\delta_{\text{m}}(\textbf{r})$. 

To measure $\textbf{g}$ for a galaxy survey, the linear relation between matter and tracer 
contrasts can be assumed, 
\begin{equation}\label{deltarelation}
\delta_{\text{tr}} = b \, \delta_{\text{m}} \, ,
\end{equation}
where $b$ is the linear bias, and we consider the LG barycentre as the origin of 
coordinates, that is $\textbf{r} = \textbf{0}$. 
If the survey volume is large enough to ensure the convergence of the 
integral~\citep{Scaramella}, one can rewrite equation (\ref{relationvelocitygravity}) as 
\begin{equation}\label{nwmethod}
\textbf{v}_{\,\text{LG}}(r) = \dfrac{H_{0}\beta}{4\, \pi\, \bar{n}} \, 
\textbf{D}_{\text{LG}}(r) \, ,
\end{equation} 
where the LG clustering dipole, $\textbf{D}_{\text{LG}}$, is defined by 
\begin{equation}\label{dipole}
\textbf{D}_{\text{LG}}(r) \equiv 
\sum_{i}^{N(r)} \dfrac{\hat{\textbf{r}}_{i}}{\phi(r_{i})\, r_{i}^{2}} \, ,
\end{equation}
summing over all objects inside the sphere of radius $r$. Equation (\ref{nwmethod}) 
is known as the \textit{number-weighted method} \citep{Erdogdu}.

\subsection{Error estimation}\label{sec3.2}

In order to correctly obtain $\beta$, several effects that can influence the gravitational 
dipole magnitude must be taken into account~\citep{Schmoldt}. 
As we already mentioned, one of them is the RSD effect, that can be avoided here because 
the distances of our ALFALFA sample are provided in real space.  
On the other hand, we must also evaluate the error introduced by shot-noise and the impact 
of using an incomplete sky coverage. 
Following, we discuss the methodology used to take into account each of them.

\subsubsection{Shot-Noise} \label{sec:shot-noise}

Following \cite{Basilakos98}, we use a set of $N_{\mbox{\sc\tiny MC}} = 144$ Monte Carlo 
realizations to calculate the error due to the shot-noise. 
This number of Monte Carlos is to equalize the number of log-normal simulations, 
$N = 144$, used to correct the dipole velocity due to the partial sky coverage of our data 
sample (see Section~\ref{partialskycoverage} for details).

The methodology to produce each realization is to randomize the angular coordinates of 
the sources, i.e., taking their right ascension and declination from uniform random 
distributions limited by the ALFALFA footprint, repositioning them inside the same region, 
while their distances remain the same, keeping unchanged the selection function. 
From the set of Monte Carlo realizations, one can calculate the covariance matrix 
due to the shot-noise as 
\begin{eqnarray}\label{shot-noise}
C_{\text{SN}}^{ij} &=& \frac{H_{0}}{4\pi\bar{n}}\frac{1}{N-1}\sum_{k=1}^{N}\, 
[D^{k}_{\text{LG}}(r_{i})-\langle D_{\text{LG}}\rangle(r_{i})] \times \nonumber \\
 &\,\,& [D^{k}_{\text{LG}}(r_{j})-\langle D_{\text{LG}}\rangle(r_{j})],
\end{eqnarray}
where $D_{\text{LG}}^{k}\equiv|\textbf{D}_{\text{LG}}^{k}|$ 
is the dipole magnitude calculated for the $k\:\!$th Monte Carlo realization and 
$\langle D_{\text{LG}}\rangle$ is the average dipole over all the 
$N_{\mbox{\sc\tiny MC}} = 144$ Monte Carlo realizations.

\subsubsection{Correction procedure for incomplete sky coverage}\label{partialskycoverage}

The incomplete sky coverage of the sample in analysis, shown in Fig. \ref{fig:skyplot}, 
certainly bias the measurement of direction and magnitude of the 
LG dipole velocity, which must be corrected accordingly. 
The idea is to use simulated catalogues to perform both, full and partial sky dipole analyses, that allow to find the corrected LG dipole velocity.

Consider a set of $N$ full-sky (FS) log-normal simulated catalogues, 
to which we apply the ALFALFA footprint to obtain the respective partial sky (PS) log-normal 
simulated catalogues.
Then, we measure the dipoles of the $i\:\!$th catalogue,
$\textbf{D}^{\text{FS},\,i}_{\text{sim}}$ and $\textbf{D}^{\text{PS},\,i}_{\text{sim}}$, 
for $i=1,\cdots,N$, for the FS and PS cases, respectively. 
The difference, $\textbf{X}^{i}(r)$, defined as 
\begin{equation}\label{correctionterm}
\textbf{\textbf{X}}^{i}(r) \equiv \textbf{D}^{\text{FS},\,i}_{\text{sim}}(r) - 
\textbf{D}^{\text{PS},\,i}_{\text{sim}}(r), 
\end{equation}
is used to correct our partial sky LG dipole velocity measurement. 
Now we construct the corrected LG dipole velocity as
\begin{equation}\label{LGcorrection}
\textbf{v}_{\text{LG}}(r)\beta^{-1} \equiv 
\frac{H_{0}}{4\pi\bar{n}}\frac{1}{N}\sum_{i=1}^{N} \,\left[ \textbf{D}^{\text{PS}}_{\text{HI}}(r) 
+ \textbf{X}^{i}(r) \right] \,,
\end{equation}
correcting both direction and module of the LG dipole velocity for a PS data 
catalogue, and, for this, appropriate for the ALFALFA catalogue 
(see Appendix~\ref{apendiceC} for details of this correction procedure).

To obtain a good performance with this correction procedure, one has to restrict the set of FS log-normal simulations to those whose clustering dipole direction are close to the LG velocity direction in CMB frame, $(l, b) = (273^{\circ} \pm 3^{\circ}, 29^{\circ}\pm3^{\circ})$, and for this we consider those maps where the misalignment between both directions is less than $30^{\circ}$\footnote{The robustness of our results has been tested for several values of maximum misalignment: $20^{\circ}, 30^{\circ}, 40^{\circ}$, achieving basically the same results for each case.}. The ideal situation would be to consider only simulations where this value is $0^{\circ}$, but this clearly would be a fine tuning, and scientifically invalid, approach. By doing this, we are left with 144 catalogues from a total of 4000 log-normal simulations produced\footnote{We tested the robustness of this, apparently small, number of catalogues. We first perform the analyses with these 144 catalogues and calculate $f\sigma_{8}$; then we add new 1000 FS log-normal catalogues to the original 4000, selecting a total of 191 catalogues, we redo the analyses and again calculate $f\sigma_{8}$ obtaining the same result.}. This is not strange, as already noticed by \cite{Kolokotronis}, to find suitable simulations that exhibit the main features of our LG universe is a difficult task, which decreases considerably the number of suitable catalogues for analyses.

This dipole correction procedure was done using a set of simulated log-normal catalogues 
constructed with the public 
code\footnote{\url{https://bitbucket.org/komatsu5147/lognormal_galaxies/src/master/}} 
presented in~\cite{Agrawal17}. 
The log-normal approach assumes that the matter and galaxy density fields can be 
represented by a log-normal probability density function. 
In \cite{Agrawal17}, the authors show the remarkable agreement between the input and 
measurements for the correlation function and power spectrum, important features used to 
analyse galaxy clustering in surveys. 

In table~\ref{table1} we show all the input parameters needed to generate our log-normal 
FS simulated catalogues. 
In the first column we present the survey configuration: the box dimensions, $L_{x}$, 
$L_{y}$, and $L_{z}$, the number of 
galaxies\footnote{This number is not constant in all realizations, but their fluctuations 
around $N = 7798$ (the number of HI line sources of the catalogue in analysis) are not 
significant.}, 
$N_g$, the redshift at which we generate the input power spectrum, $z$, and the bias, $b$. 
The matter power spectrum, $P(k)$, is calculated using Eisenstein \& Hu (EH) transfer 
function~\citep{Eisenstein98}. 
The code uses this approach by default, in case one does not provide a table of $P(k)$ 
values calculated externally (see Appendix \ref{apendiceD} for a check of the accuracy of 
the EH approach).
All these parameters were chosen in order to reproduce the ALFALFA survey. 
In the second column of table~\ref{table1} we observe the cosmological parameters, as 
given by the~\cite{Planck18}. 
For the bias choice, see Section \ref{sec3.3}.

It is worth mentioning that, even though the log-normal simulations do not reproduce 
accurately the velocity field when compared with N-body simulations, as shown 
by \cite{Agrawal17} using the linearised continuity equation of the matter fields, our 
analyses using equations (\ref{nwmethod}) and (\ref{dipole}) use only the galaxy positions. 
Therefore, as our tests suggest in appendix \ref{apendiceC}, the log-normal catalogues 
can be used to obtain robust results of the ALFALFA dipole convergence and, consequently, 
an accurate measurement of $f\sigma_{8}$.

Finally, one can calculate the error in equation (\ref{LGcorrection}) using the same procedure used to compute the shot-noise error. 
The covariance matrix due to the partial sky coverage using the log-normal 
(LN) catalogues can be calculated as
\begin{eqnarray}\label{Verror}
C_{\text{LN}}^{ij} &=& \frac{1}{N-1}\sum_{k=1}^{N}\, ([v_{\text{LG}}(r_{i})\beta^{-1}]^{k}-\langle [v_{\text{LG}}(r_{i})\beta^{-1}]\rangle)\times\nonumber\\
&\,\,& ([v_{\text{LG}}(r_{j})\beta^{-1}]^{k}-\langle [v_{\text{LG}}(r_{j})\beta^{-1}]\rangle),
\end{eqnarray}
where $v_{\text{LG}} \equiv |\textbf{v}_{\text{LG}}|$, $[v_{\text{LG}}(r)\beta^{-1}]^{k}$ 
is the LG velocity for the $k\,$th log-normal realization, and 
$\langle [v_{\text{LG}}(r)\beta^{-1}]\rangle$ is the average LG velocity 
over $N$ realizations.

Notice that to obtain the errors in both $\beta$ and $f\sigma_{8}$, we combine both 
covariance matrices, that is,
\begin{equation}\label{fullmatriz}
C_{\text{SN+LN}}^{ij} = C_{\text{SN}}^{ij} + C_{\text{LN}}^{ij}.
\end{equation}

\begin{table}
\caption{Survey configuration and cosmological parameters from the Planck last data 
release~\citep{Planck18} used to generate the set of $N_{\mbox{s}} = 4000$ log-normal 
realizations used in the analyses.}
\centering
\setlength{\extrarowheight}{0.2cm}
	\begin{tabular}{c|c}
	\hline
	Survey configuration          & Cosmological parameters         \\ 
\hline
	$z=0.0$                              & $\Omega_{c}h^{2}=0.1202$       \\  
	$b=1.0$                              & $\Sigma m_{\nu}=0.0600$         \\
	$N_{g}=2\times 10^{5}$     & $n_{s}=0.9649$                         \\
	$L_{x}=230$                     & $\ln(10 A_{s})=3.045$                  \\
	$L_{y}=230$                     & $\Omega_{b}h^{2}=0.02236$       \\
	$L_{z}=230$                     & $h=0.6727$       \\ \hline                          
	\end{tabular}
\label{table1}
\end{table}

\subsection{Measuring $\sigma_{8}^{\text{tr}}$}\label{sec3.3}

In order to perform a measurement of $f \sigma_{8}$ in the local Universe using our 
data sample, we first calculate $\beta$, using equation (\ref{nwmethod}), and, then, 
$\sigma_{8}^{\text{tr}}$, the matter fluctuation of the HI line sources of our 
data sample in spheres of 8 Mpc\,/$h$. 
For galaxy samples, $\sigma_{8}^{\text{tr}}$ was observed to be close to 
1~\citep{Juszkiewicz,Boruah}. However, this result will depend on the 
bias of the tracer. 
In the $\Lambda$CDM context one expects $\sigma_{8}^{\text{tr}} < 1$ for $b\lesssim1$.

In this work we calculate $\sigma_{8}^{\text{tr}}$ using the relationship 
\begin{equation}\label{sig8tr}
\sigma_{8}^{\text{tr}} = b_{\text{HI}} \,\sigma_{8} \,,
\end{equation}
where $\sigma_{8} = 0.8120 \pm 0.0073$ 
from~\cite{Planck18}\footnote{
This value of $\sigma_{8}$ corresponds to $z=0$; 
assuming the $\Lambda$CDM fiducial model of table~\ref{table1} one finds: 
$\sigma_{8}(z=0) - \sigma_{8}(z=0.013) = 0.0056$, a difference 
smaller than the error bar that does not modify our main result. 
For this, we use $\sigma_{8}(z=0.013) \simeq \sigma_{8}(z=0) = 0.8120 \pm 0.0073.$}. 
The linear bias, $b_{\text{HI}}$, for the HI line sources can be obtained from the work 
of~\cite{Martin12}, where they calculated the bias for different scales. 
In Fig. 10 of this reference we observe, in the interval 3-30 Mpc\,/$h$ , 
a fluctuation around 1, where the HI tracer reflects the underlying matter distribution. 
Then, we take the data points in this scale range and fit a horizontal line to them, obtaining 
\begin{equation}\label{bias}
b_{\text{HI}} = 0.99 \pm 0.11 \, .
\end{equation}
This bias value motivated us to fix $b = 1.0$ to generate the log-normal simulations. 
Using equation (\ref{sig8tr}), we obtain the variance for our sample, that is, 
$\sigma_{8}^{\text{tr}} = 0.80 \pm 0.09$.

\section{Results}\label{sec4}

\begin{figure}
	\centering
	\includegraphics[width=8.0cm,height=5.0cm]{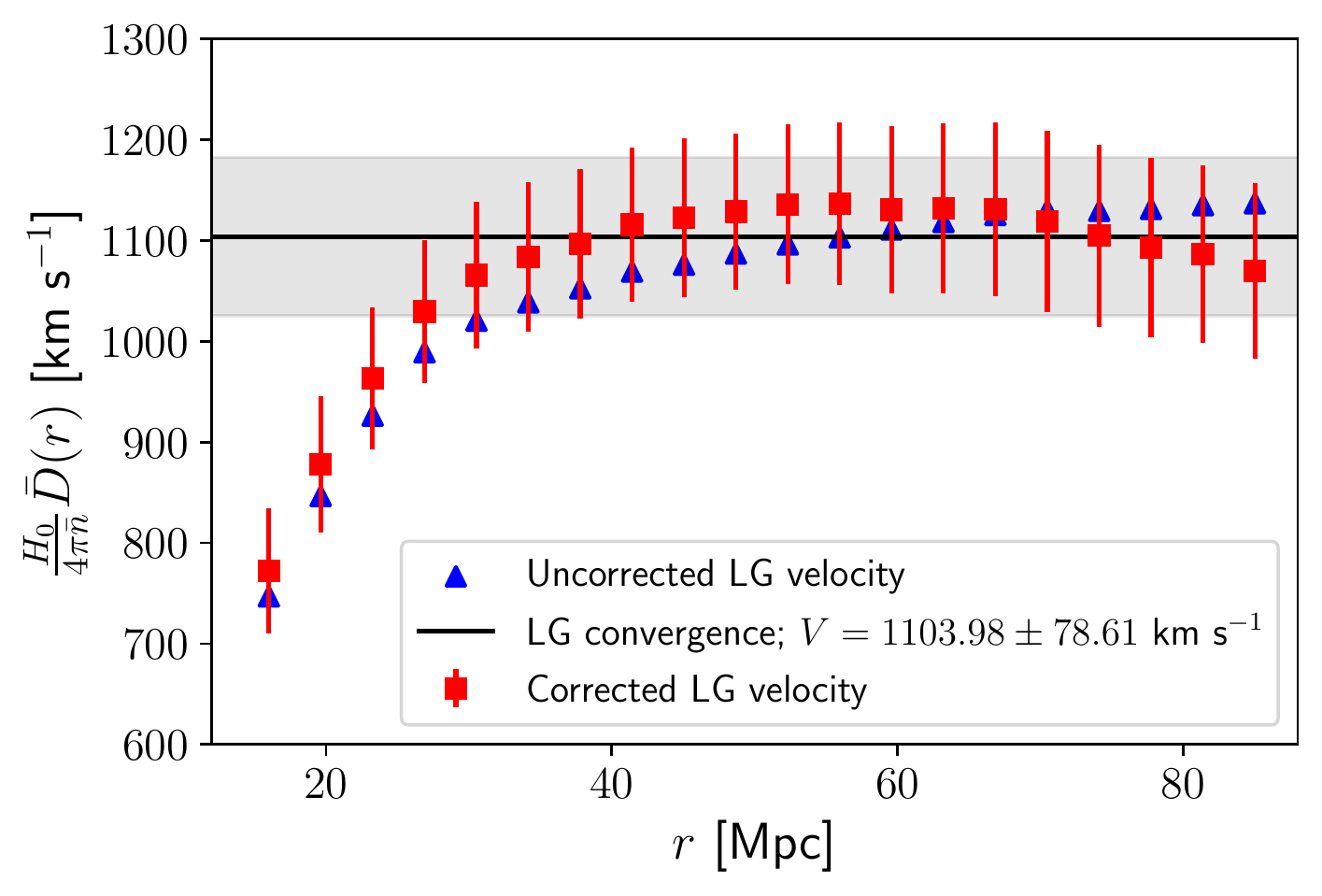}
	\caption{The ALFALFA LG velocity function, in real space, as a function of the 
		radial distance from the observer to a surface of radius $r$, where the dipole is evaluated. 
		The error bars take into account cosmic variance and shot-noise errors (see the text for 
		details).
		The convergence value, $V$, and its 1$\sigma$ error are represented by the black 
		horizontal line and the shaded region, respectively.
	}
	\label{fig:alfalfadipole}
\end{figure}

\begin{figure}
	\centering
	\includegraphics[width=8.0cm,height=6.0cm]{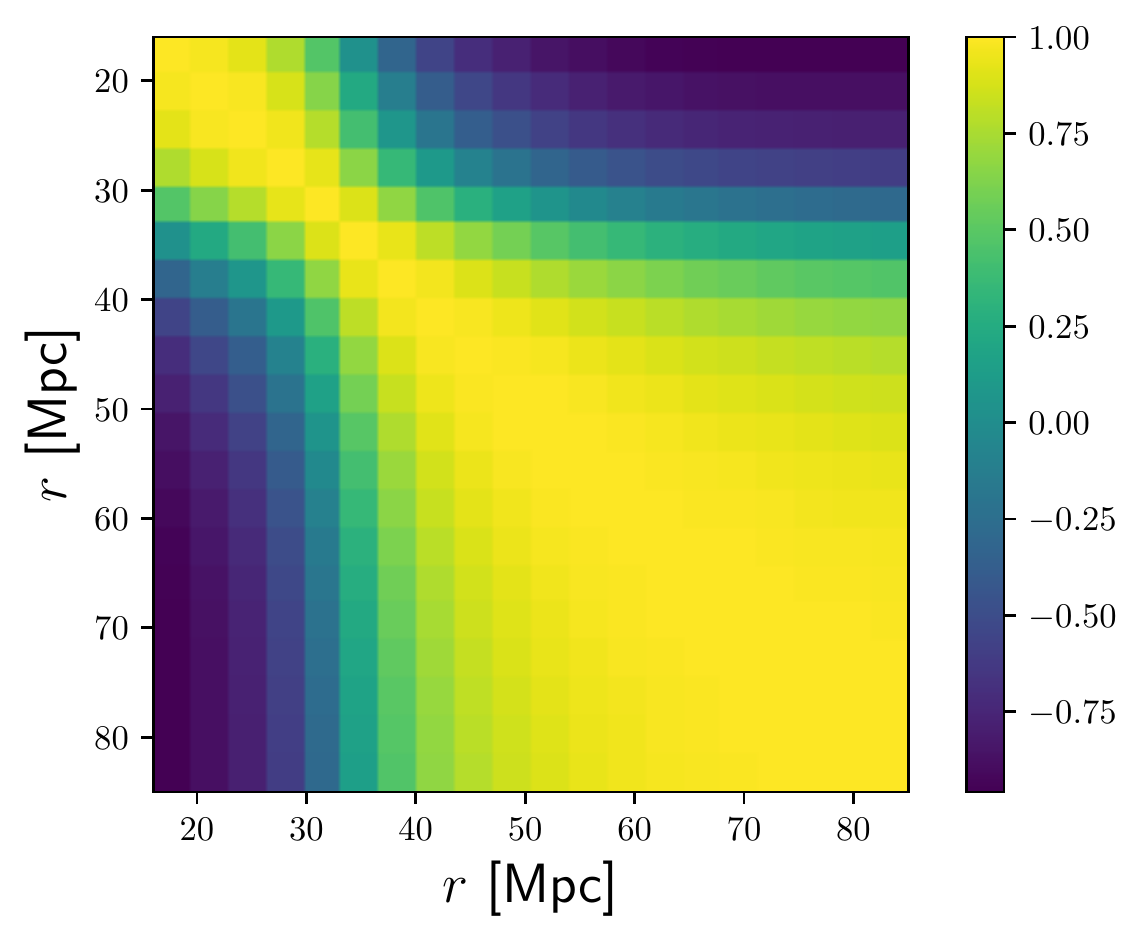}
	\caption{
		Reduced covariance matrix, $C^{ij} / \sqrt{C^{ii}\times C^{jj}}$, obtained from equation (\ref{fullmatriz}). 
		The covariance matrix is obtained combining the shot-noise error, using Monte Carlo 
		realizations, and the error due to the correction for partial sky survey, using log-normal 
		simulations.
	}
	\label{fig:covmatrix}
\end{figure}

In Fig. \ref{fig:alfalfadipole} we show the result of 
applying equations (\ref{nwmethod}) and (\ref{LGcorrection}) to 
our ALFALFA sample, where $ \bar{n} = 0.04 $ Mpc$ ^{-3} $ and 
$H_{0} = 67.27$ km/s/Mpc \citep{Planck18}. 
The error bars come from the diagonal terms of the covariance matrix given by equation (\ref{fullmatriz}) (see Fig. \ref{fig:covmatrix}), accounting 
for the cosmic variance, that comes from the 144 log-normal 
simulations, using equation (\ref{Verror}), and the shot-noise, from the 144 Monte Carlo 
realizations using equation (\ref{shot-noise}). 
In the same plot, we compare the corrected (red squares) and the uncorrected (blue 
triangles) LG velocity. One can see that, in average, for scales smaller than $\sim 70$ Mpc 
the PS uncorrected analysis underestimate the dipole amplitude, with an opposite 
behaviour for larger scales. 
The black line and the shaded region represent the convergence value and $1\sigma$ uncertainty of the LG velocity function, respectively. We describe below how we obtain this value.

\begin{figure}
\centering
\includegraphics[width=8.0cm,height=5.0cm]{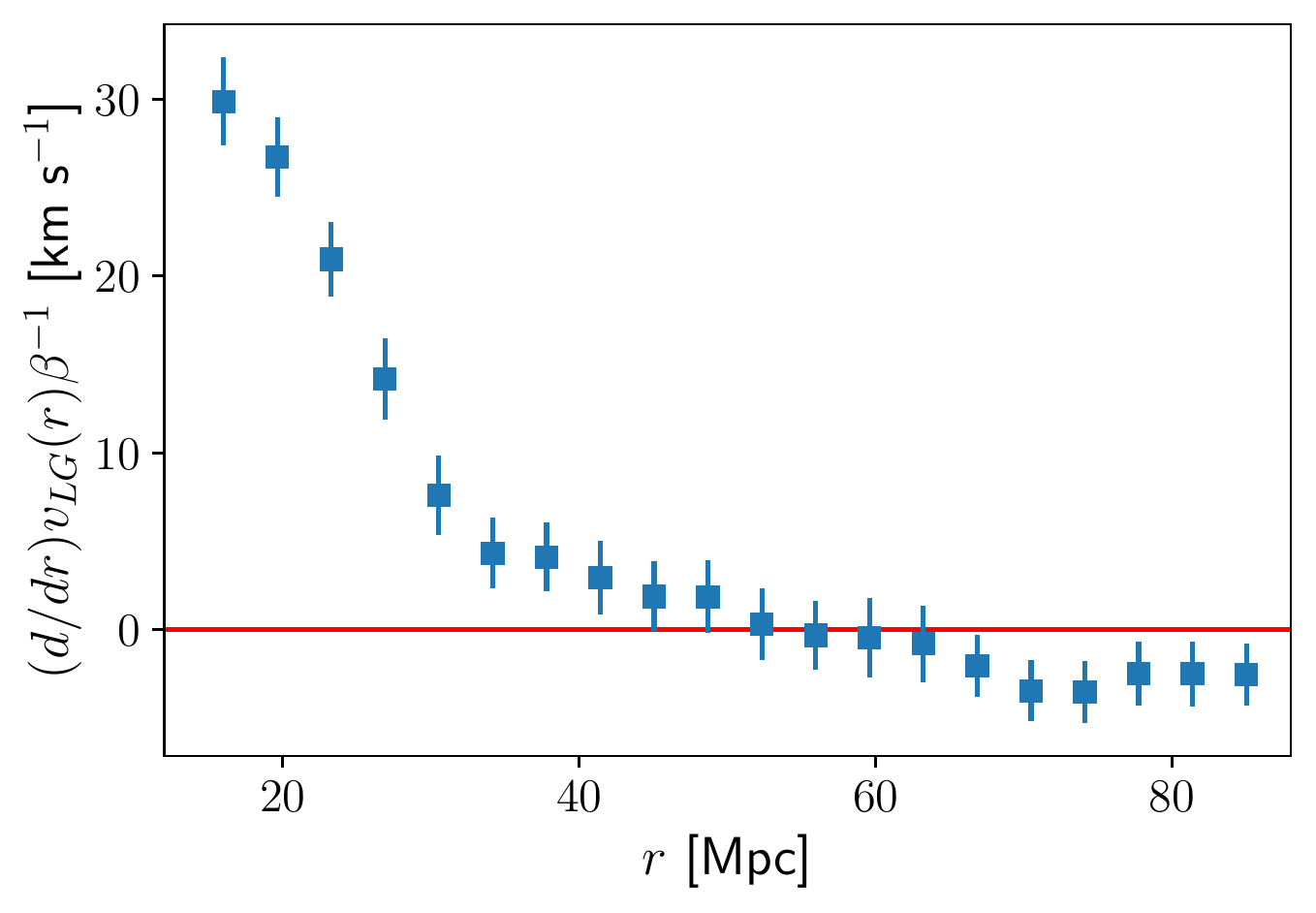}
\caption{Numerical derivative of the ALFALFA LG velocity function 
(red squares in Fig. \ref{fig:alfalfadipole}) with respect to the radial distance $r$. 
The error bars correspond to the 1$\sigma$ dispersion from the 144 log-normal 
simulations.	
}
\label{fig:derivative}
\end{figure}

To obtain the convergence value, $V$, taking into account the error bars, 
we performed a numerical derivative of the LG velocity function, that is,
\begin{equation}\label{derivativeLGcorrection}
\frac{d}{dr}\textbf{v}_{\text{LG}}(r)\beta^{-1} = 
\frac{H_{0}}{4\pi\bar{n}}\frac{1}{N}\sum_{i=1}^{N} \,\frac{d}{dr}\left[ \textbf{D}^{\text{PS}}_{\text{HI}}(r) 
+ \textbf{X}^{i}(r) \right] \,,
\end{equation}
looking for the scale interval consistent with zero, i.e., where this function attains a 
maximum, ensuring an accurate measurement of this convergence value. 
In Fig. \ref{fig:derivative} we show the result of equation (\ref{derivativeLGcorrection}) 
applied to the ALFALFA LG velocity, the 1$\sigma$ error bars are estimated performing the 
same sequence of analyses over each of the 144 simulations. 
We observe that the derivative is consistent with zero, within a $1\sigma$ confidence level, 
in the interval $45-63$ Mpc. 
Thus, to obtain the convergence value, we take the LG velocity data points in this interval 
and fit for a horizontal line to them, obtaining 
\begin{equation}\label{Vresult}
V = 1103.98 \pm 78.61~\text{km/s} \,.
\end{equation}

In the Fig. \ref{fig:alfalfadipole}, we observe that the ALFALFA corrected LG velocity is 
consistent with this convergence value (black vertical line), within the $1\sigma$ level, 
until $85$ Mpc, i.e., the limit of our analysis. 
As our selection function is of order $\phi(r = 85~\text{Mpc}) \simeq 0.3$, we are confident 
that our result does not indicate an artificial convergence, as discussed in~\cite{Scaramella}.

Then, one can find the velocity scale parameter $\beta$ from 
equation (\ref{nwmethod}) by using $V = 1103.98 \pm 78.61$ km\,/\,s value and the known 
LG peculiar velocity in the CMB frame 
$u_{\,\text{LG}} \equiv |\textbf{u}_{\text{LG}}| = 627 \pm 22$ km\,/\,s 
\citep{Courteau,Erdogdu} 
\begin{equation} \label{beta}
\beta = \frac{u_{\text{LG}}}{V} = 0.57 \pm 0.04 \,.
\end{equation}
Because $f = b \,\beta$, this measurement of $\beta$ combined with 
the bias $b$ of the sample in analysis, equation~(\ref{bias}), provides a measurement of 
the growth rate of structures $f$, 
\begin{equation} \label{f}
f \,=\, 0.56 \pm 0.07 \, ,
\end{equation}
at $\bar{z}=0.013$.

Our measurement of $\beta$ shows a good agreement with the value 
$\beta^{\Lambda\text{CDM}}$ expected in the $\Lambda$CDM model. 
In fact, equation~(\ref{growth rate parametrize}) with $\gamma=0.55$ and 
$\Omega_m = 0.3150$ (from table~\ref{table1}) gives 
$f^{\Lambda\text{CDM}} = 0.54$ at $\bar{z}=0.013$; then, using 
$b_{\text{HI}} = 0.99\pm0.11$ we have $\beta^{\Lambda\text{CDM}} = 0.54\pm0.06$.

\begin{figure}
\centering
\includegraphics[width=8.0cm,height=5.0cm]{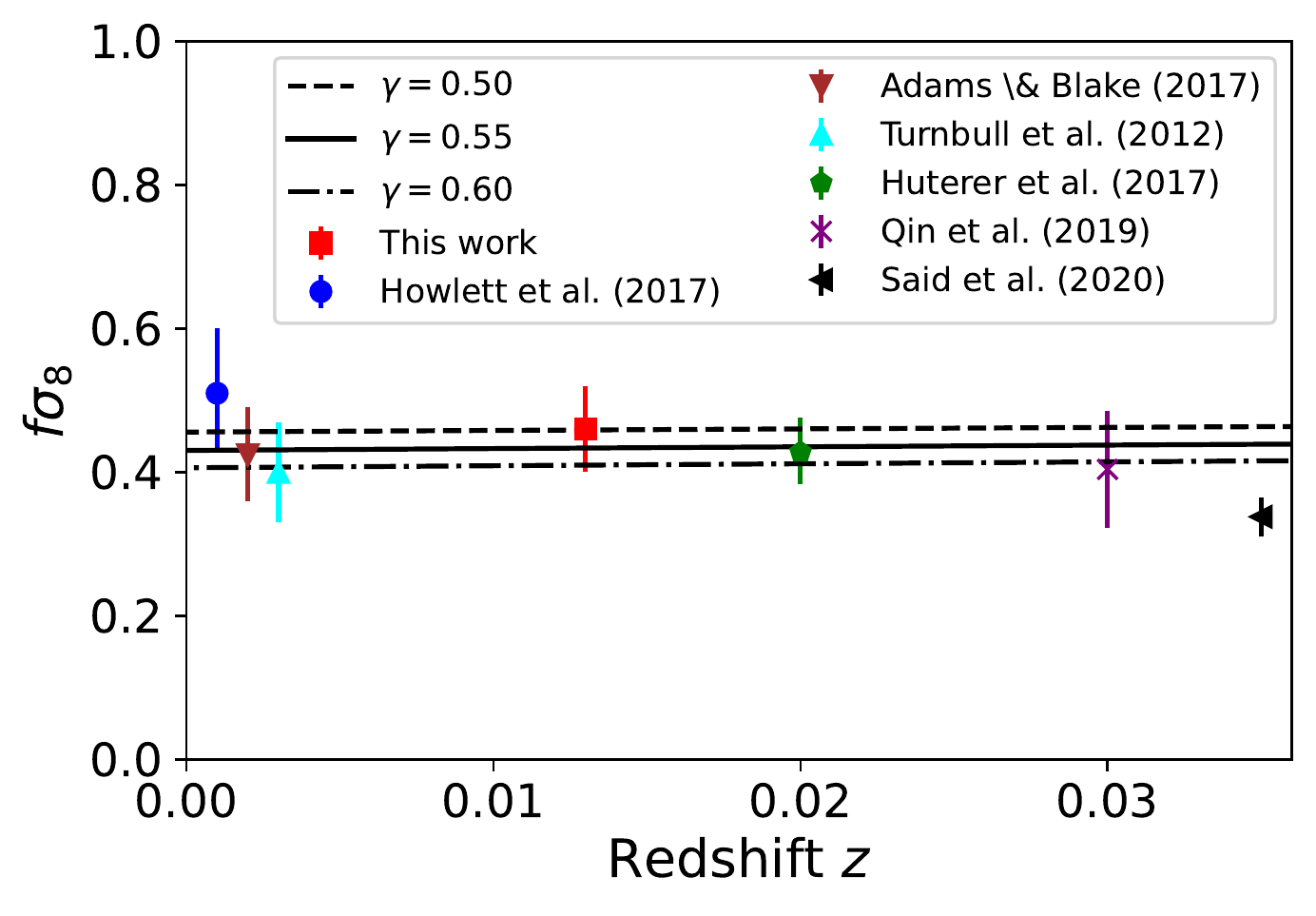}
\caption{Measurements of $f\sigma_{8}$ for the local Universe, where our 
result is shown as a red square. 
We observe a good agreement with the $\Lambda$CDM model, corresponding to the 
case $\gamma=0.55$ (see equation (\ref{growth rate parametrize})). 
The blue dot is the work of \protect\cite{Howlett17}, using the 2MTF galaxies, and 
the inverted brown triangle is the outcome of \protect\cite{Adam17} for the 6dF galaxy survey 
data. The work of \protect\cite{Turnbull}, cyan triangle, used a compilation of SNe Ia.
These three measurements are slightly shifted to the right for a better visualization, but they 
were calculated at $z \simeq 0$. The green pentagon shows the result of \protect\cite{Huterer17} combining low redshift 
SNe Ia with the 6dF galaxy survey. 
The last two points correspond to~\protect\cite{Qin19} (who combine the 2MTF and 6dF 
galaxies) and~\protect\cite{Said20} (who performed joint analyses of the 6dFGS and SDSS 
data) at $z=0.03$ and $z=0.035$, respectively.}
\label{fig:fsig8}
\end{figure}

Finally, we can obtain our main result $f\sigma_{8}$; for this we use equation~(\ref{fsig8}) 
to combine $\beta$, from equation (\ref{beta}), with $\sigma_{8}^{\text{tr}} = 0.80 \pm 0.09$, 
obtained using equations (\ref{sig8tr}) and (\ref{bias}), to get 
\begin{equation}\label{fsig8result}
f\sigma_{8} = 0.46 \pm 0.06 \,,
\end{equation}
at $\bar{z} = 0.013$, consistent with the $\Lambda$CDM model at $1\sigma$ confidence 
level, $[f\sigma_{8}]^{\Lambda\text{CDM}} = 0.43\pm 0.02$. 
In Fig. \ref{fig:fsig8} we display, for comparison, our result together with a sample of 
measurements of $f\sigma_{8}$ at low redshift, performed through diverse methodologies 
that analyse several cosmological tracers.

As a robustness test for the $H_0$ value, a parameter in the calculations of the LG dipole 
velocity, we have produced another set of log-normal simulations with the hypothesis 
$H_0 = 74.03$ km/s/Mpc~\citep{riess2019} (i.e., $h = 0.7403$). 
We have repeated the analyses finding: 
	$\beta = 0.51 \pm 0.02$ and $f \sigma_8 = 0.41 \pm 0.05$, 
	which reproduces, within $1\,\sigma$, the results already obtained. 
	This reveals that the value of the Hubble constant $H_0$ has a limited impact on our 
	analyses, and that our results are robust under different values of $H_0$ reported in the 
	literature.

\section{Conclusions}\label{sec5}

The structures growth data of the observed Universe has the potential to differentiate 
between the metric theory that supports the standard cosmological model, that is GR 
theory, from those  based on modified gravity models. 
Due to this scenario, efforts are being done to analyse several cosmological tracers with 
diverse approaches and methodologies. 
One of these, is the gravitational dipole technique~\citep{Hudson93,Scaramella}. 
In summary, this methodology compares the peculiar velocity of the LG of galaxies, inferred 
from the CMB dipole, to the LG gravitational acceleration calculated from a given 
cosmological tracer. 
Using the catalogue of extra-galactic HI line sources provided by the ALFALFA 
survey~\citep{Haynes18}, we investigate the growth rate of cosmic structures in the local 
Universe.

In fact, these analyses can be biased by various systematic effects, such as shot-noise, 
RSD, and non-linear effects, as well as the incomplete sky coverage of the survey. 
As discussed in Section~\ref{sec3.2}, the latter is the main source of systematics in our 
analyses. 
To correct the LG velocity dipole for the effect produced by the fact that the observed 
data in the ALFALFA survey covers a partial region of the celestial sphere, instead of the full 
sky, we use sets of FS and PS log-normal simulations according to the correction procedure 
described in Section~\ref{partialskycoverage}. 
In contrast, while the RSD effect can be avoided, since our data sample allows the dipole 
calculation in real space, the shot-noise contributes with a relatively small error due to the 
high number density of the sample, as shown in the analyses performed with the set of 
Monte Carlo realizations.

Additionally, our analyses show that it is possible to estimate the product of the growth rate 
and the matter fluctuation, $f\sigma_{8}$, through the gravitational dipole 
approach~\citep{Strauss,Scaramella} using a partial sky catalogue, 
as long as the bias of the cosmological tracer and the correction due to the partial sky survey 
are carefully taken into account. 
This way, we found that the magnitude of the dipole velocity calculated from the ALFALFA 
sample reaches the convergence around $60$ Mpc, and its magnitude leads to our estimate 
of the velocity scale parameter, $\beta = 0.57 \pm 0.04$.
Together with our measurement of the matter fluctuation in the local Universe, 
$\sigma_{8}^{\text{tr}} = 0.80 \pm 0.09$, it provides the value 
$f \sigma_{8} = 0.46 \pm 0.06$ at $\bar{z} = 0.013$.
This measurement is in good agreement, at $1\sigma$ level, with the value obtained 
in the $\Lambda$CDM concordance model: 
$[f\sigma_{8}]^{\Lambda\text{CDM}} = 0.43\pm 0.02$. 
As observed in Fig. \ref{fig:fsig8}, where we show a small compilation of $f \,\sigma_{8}$ 
values in the local Universe, our result is in good consonance with measurements obtained 
analysing several cosmological tracers through methodologies different from ours. 
Furthermore, our analyses of the ALFALFA sample also provide a 
measurement of the growth rate of structures $f = 0.56 \pm 0.07$, at $\bar{z}=0.013$.


\section*{Acknowledgements}
The authors thank PROPG-CAPES/FAPEAM program, CNPq, CAPES, and FAPESP 
(process no. 2019/06040-0) for the grants under which this work was carried out. 

\section*{Data Availability}
The data underlying this article were accessed from \url{http://egg.astro.cornell.edu/alfalfa/data/index.php}. The derived data generated in this research will be shared on reasonable request to the corresponding author.

\appendix

\section{Testing the impact of distance uncertainties on dipole analyses}\label{apendiceA}

In this appendix we show that the possible underestimation in the measurements 
of the distance errors in the ALFALFA catalogue produces a negligible effect in our dipole 
analyses. 
To show this we perform the following test. 
Consider our ALFALFA data set of $N = 7798$ distance values: 
$\{ d_{i} \}, \, i=1,2, \cdots, 7798$. 
We generate 4000 Monte Carlo realizations, where each realization contains $7798$ 
simulated distance values, the $i\:\!$th distance $d_{i}^{sim}$ is taken from a normal 
distribution with mean value $d_{i}$ (the true value) and standard deviation 
$\sigma_i = 0.2\,d_{i}$. 
That is, we consider Monte Carlo realizations as simulated catalogues with `wrong' 
distance values, quantities that deviates in $20\%$ on average from the original `true' 
values given in the ALFALFA catalogue, a conservative deviation as suggested by the 
information contained in \cite{Boruah}.

\begin{figure}
	\centering
	\includegraphics[width=8.0cm,height=5.0cm]{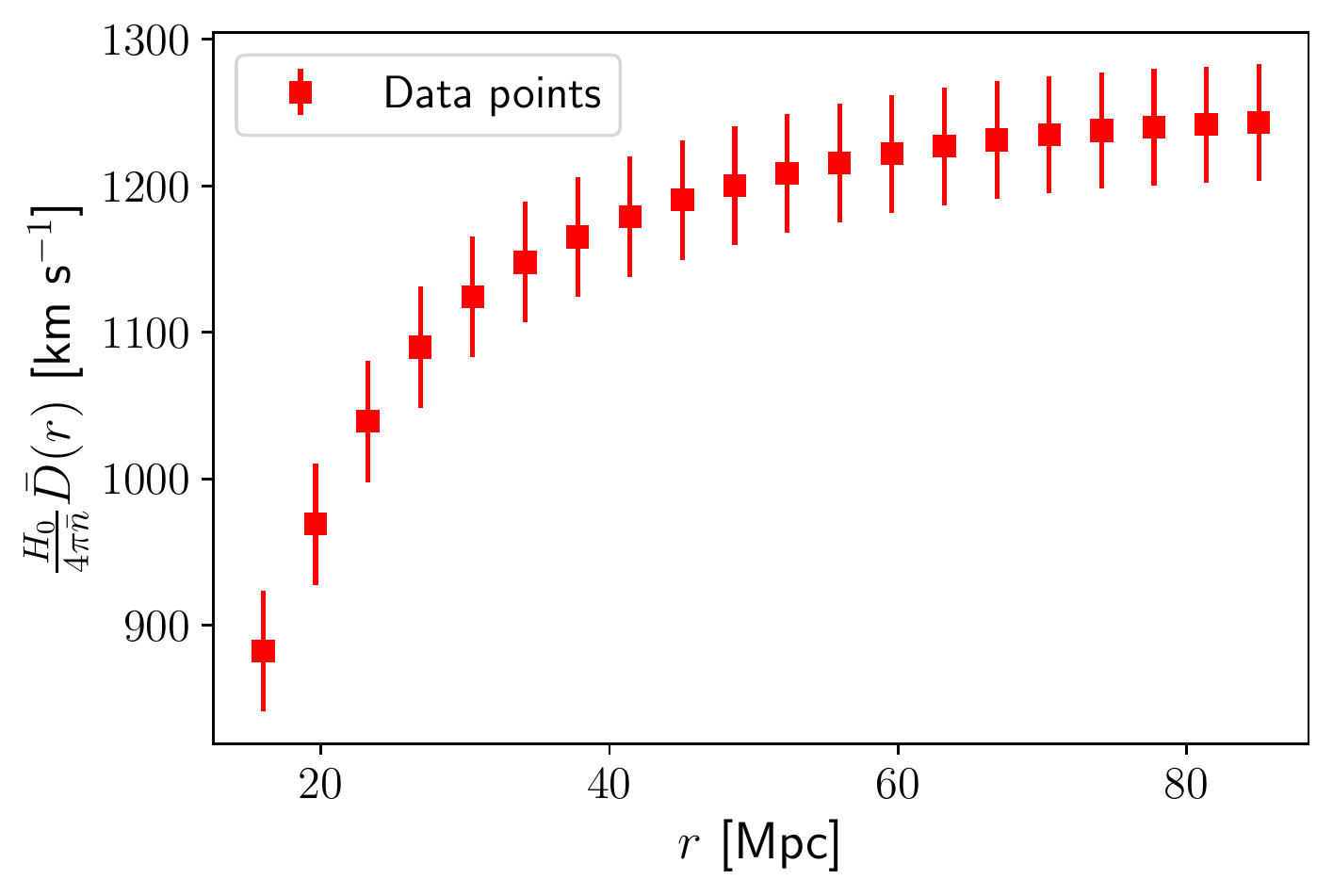}
	\includegraphics[width=8.0cm,height=5.0cm]{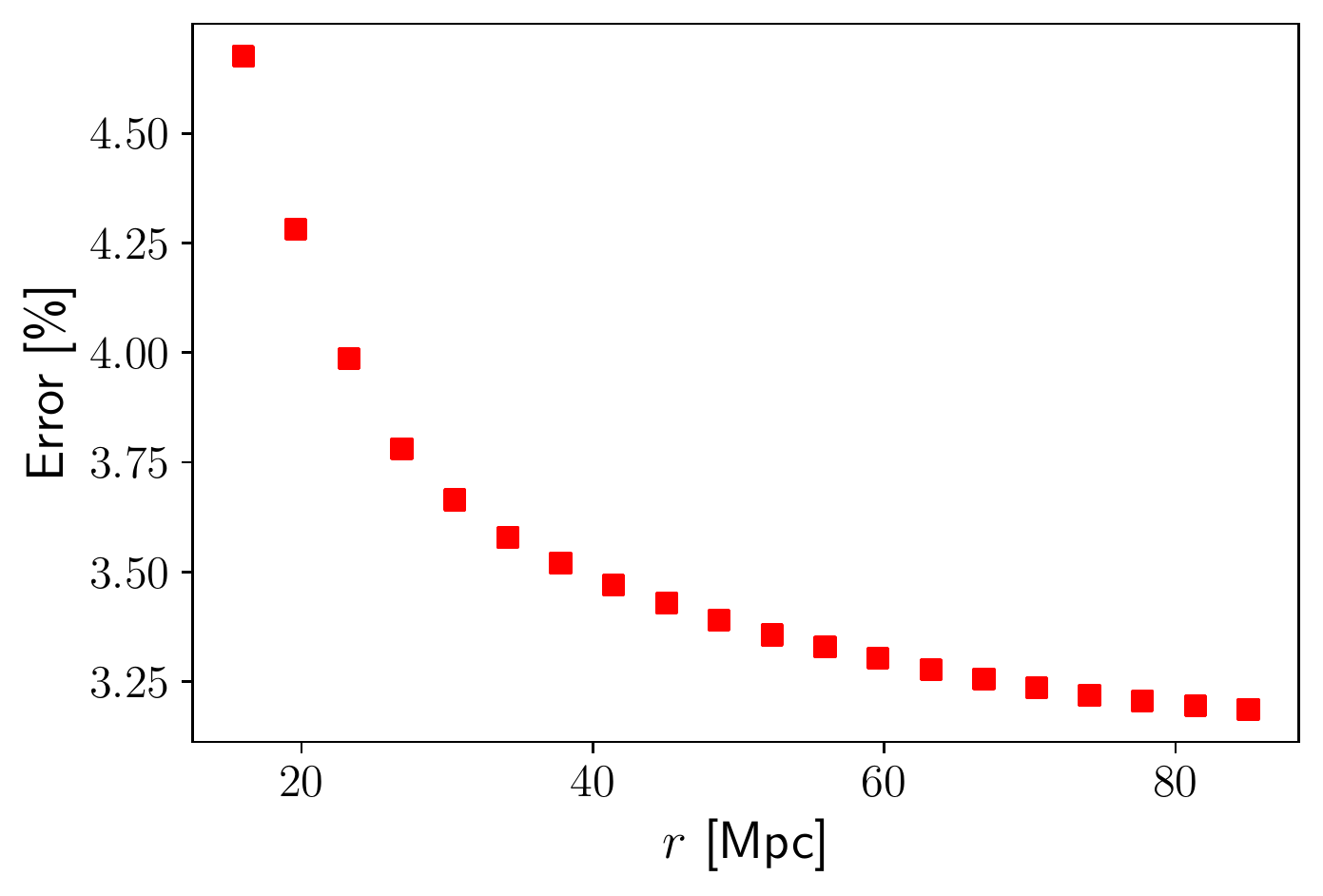}
	\caption{Upper panel: Dipole analyses of the ALFALFA catalogue with error 
		bars given by 4000 Monte Carlo realizations generated imposing large artificial errors in 
		distance measurements. 
		Lower panel: Error, in percentage, with respect to the dipole amplitude.
		See the text for details.}
	\label{fig:dipolemc}
\end{figure}

In the upper panel of Fig. \ref{fig:dipolemc}, we show the result of this test. 
The squares represent the data values obtained analysing the dipole value for the 
ALFALFA catalogue, and the error bars correspond to the standard deviation for the same 
analysis done with each one of the 4000 Monte Carlo realizations. 
According to this test, the assumed deviations in the distance values have a negligible 
effect on the determination of the dipole, as observed in the lower panel of 
Fig. \ref{fig:dipolemc}, where one can see that the error bars correspond to 
less than 5\% of the measured dipole amplitude. 

\section{Test for peculiar velocities}\label{apendiceB}

\begin{figure}
	\centering
	\includegraphics[width=8.0cm,height=5.0cm]{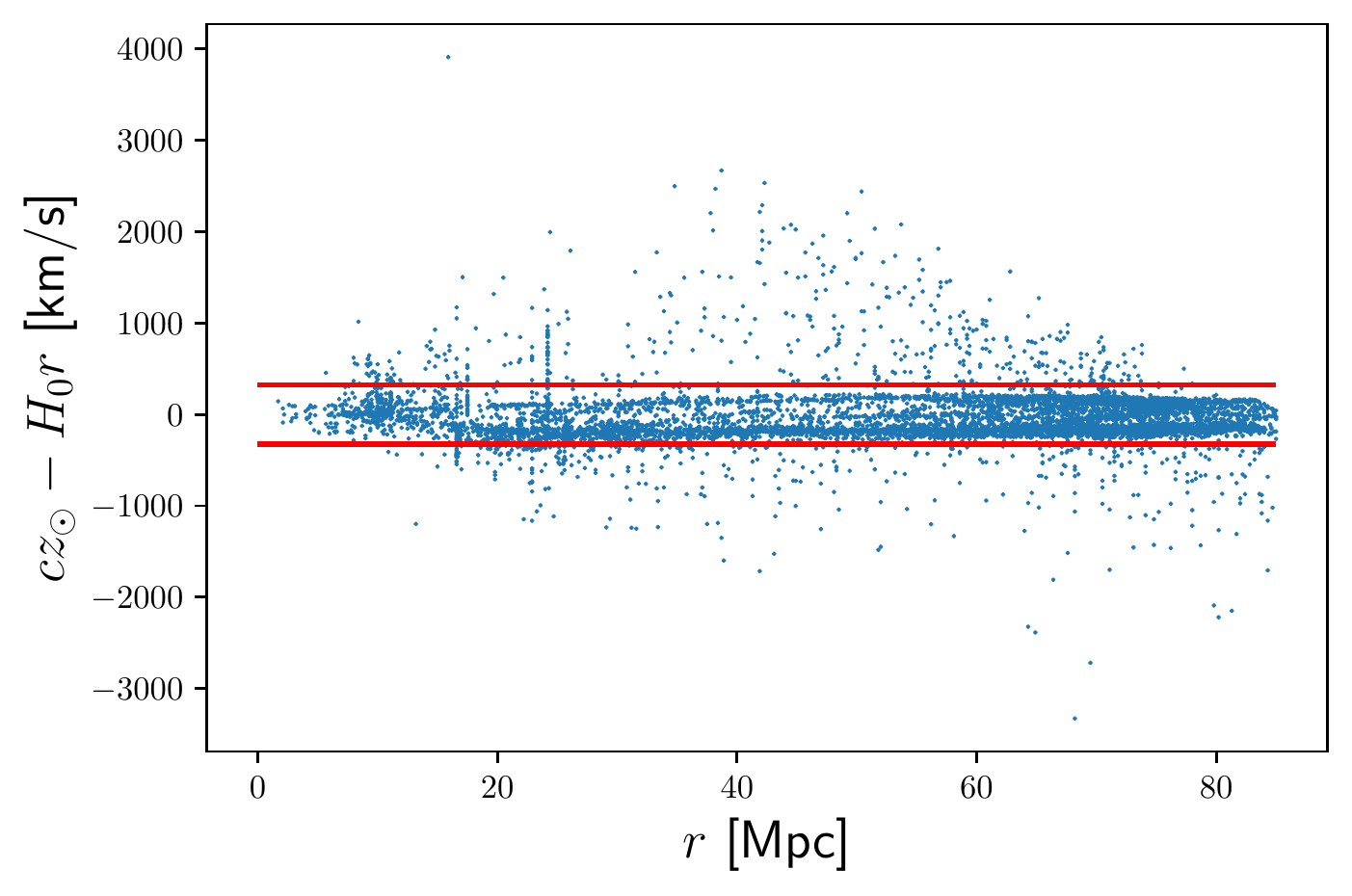}
	\includegraphics[width=8.0cm,height=5.0cm]{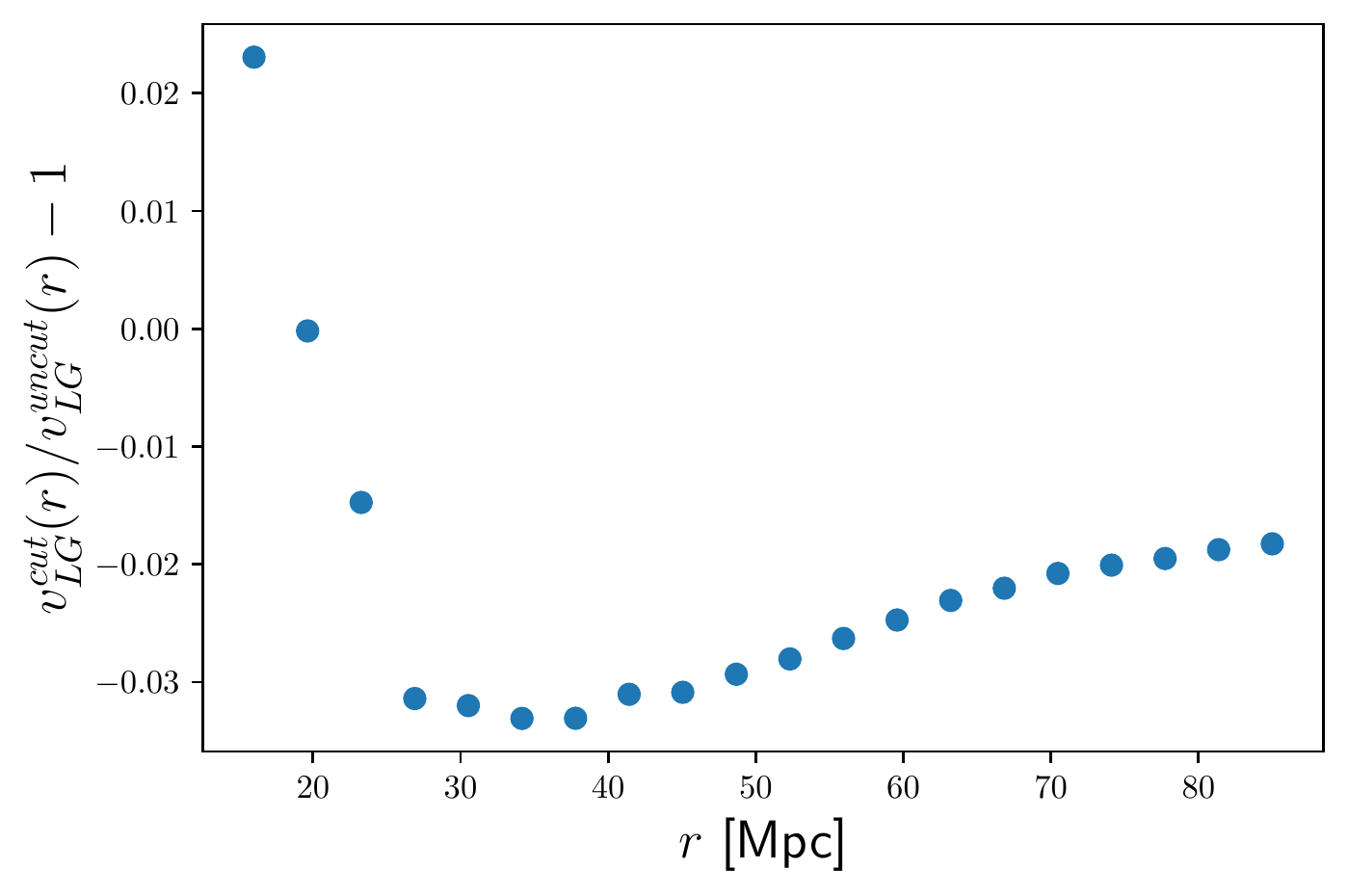}
	\caption{Upper panel: Velocities of the ALFALFA HI line sources with 
		respect to the Hubble flow within (i.e., inside the red lines) and out (i.e., outside the red 
		lines) $1\sigma$ dispersion of the Hubble flow. 
		Lower panel: Relative difference of the uncorrected LG velocity of the test sample, 
		obtained removing the cosmic objects with large peculiar velocities ($v_{LG}^{\text{\it cut}}$), 
		with respect to the uncorrected LG velocity of the original sample 
		($v_{LG}^{\text{\it uncut}}$).}
	\label{fig:velocitydispersion}
\end{figure}

From Fig.~\ref{fig:cut85} one observes that some HI line sources deviate largely from 
the Hubble flow, indicating that these objects undergo strong gravitational interactions in 
the local Universe, as a consequence they have large peculiar velocities. 
To evaluate if a set of large peculiar velocities can affect our dipole measurement we 
performed a test. 
We calculate, for this set of $N = 7798$ HI line sources, the $1 \sigma$ dispersion of 
velocities compared with the velocity expected in the Hubble flow, obtaining $325$ km/s, 
represented by red lines in the upper panel of Fig.~\ref{fig:velocitydispersion}. 
After that, we remove from the sample those objects with velocities out of this 1$\sigma$ 
dispersion level, remaining a test sample of $N^{test} = 6881$ HI line sources. 
We then compute the LG velocities for each of these two samples: with $N = 7798$ and 
with $N^{test} = 6881$ objects.
In the lower panel of Fig.~\ref{fig:velocitydispersion} we compute the relative difference between these uncorrected LG velocities, which shows a maximum deviation of $\sim 3\%$. 
We further investigate the impact of these objects with large peculiar velocities in the 
measurement of the corrected LG velocity. 
Our result shows that these uncorrected LG velocities are within the $1 \sigma$ error of 
the corrected LG velocity. 
Therefore, we conclude that the effect caused by the peculiar velocities in our analyses 
is negligible, and their impact is within the error of our dipole measurement.

\section{The correction procedure of the LG dipole velocity}\label{apendiceC}

With the procedure used in section \ref{partialskycoverage} we correct the partial sky clustering dipole measurement, 
$\textbf{D}^{\text{PS}}_{\text{data}}(r)$, as 
\begin{equation}\label{correct}
\textbf{D}^{\text{corrected},\,i}_{\text{data}}(r) 
= \textbf{D}^{\text{PS}}_{\text{data}}(r) + \textbf{X}^{i}(r) \,,
\end{equation}
$i=1,2,\cdots,N$, where $N$ is the number of simulated catalogues used in the correction 
analyses. 
The correction term, $\textbf{X}^{i}(r)$ defined in equation~(\ref{correctionterm}), 
is the vectorial difference between the full sky (FS) and partial sky (PS) clustering dipoles 
obtained from a set of $N$ log-normal simulations, for $i = 1, \cdots, N$. 
In equation~(\ref{correct}), the vector $\textbf{X}^{i}(r)$ contains the information loosed due 
to the partial sky coverage of the data survey. 
Thus, this procedure produces, for each radial distance $r$, a set of $N$ values 
$\{ \text{D}^{\text{corrected}, \,i} \} = \{ | \textbf{D}^{\text{corrected}, \,i} | \}$ that we use to 
find the correct clustering dipole. 
The corrected clustering LG dipole $\overline{\text{D}}^{\text{\,corrected}}(r)$ is the 
average of the set of $N$ values: $\{ \text{D}^{\text{corrected}, \,i}(r) \}$, and the associated 
error is the standard deviation of this set. 
These data, $\frac{H_{0}}{4\pi\bar{n}} \, \overline{\text{D}}^{\text{\,corrected}}(r)$, are 
plotted as red squares in Fig.~\ref{fig:alfalfadipole}, 
while the blue triangles correspond to the partial sky uncorrected data 
$\frac{H_{0}}{4\pi\bar{n}} \, \text{D}^{\text{PS}}_{\text{data}}(r)$.

Finally, the corrected clustering dipole, $\overline{\text{D}}^{\text{\,corrected}}(r)$, is  related 
to the LG velocity, $\text{v}_{\text{LG}}(r)$, through $\beta$ as 
\begin{equation}
\text{v}_{\text{LG}}(r) \,\beta^{-1} = 
\frac{H_{0}}{4\pi\bar{n}} \, \overline{\text{D}}^{\text{\,corrected}}(r) \, ,
\end{equation}
equivalent to the equation~(\ref{LGcorrection}).

A robustness test is in due here, to show the performance of this correction procedure. 
Firstly, we select the set of $N$ FS log-normal simulated catalogues where the misalignment 
of their clustering dipole and the CMB dipole direction is less than $30^{\circ}$. 
Then we have two sets of simulated catalogues: $N$ FS log-normal maps, and 
$N$ PS log-normal maps (obtained from the first set after applying the ALFALFA footprint). 
From this set of PS catalogues, we select one of them to be considered the {\em data} 
catalogue. 
The remaining $N-1$ PS catalogues and the $N-1$ FS catalogues will be used in the 
correction procedure of this {\em data} catalogue.

The second step is to calculate the uncorrected and the corrected clustering dipoles of this 
{\em data} catalogue according to equations~(\ref{correctionterm}) and~(\ref{correct}). 
After that, we repeat these calculations using each one of the other $N-1$ simulations as 
the {\em data} catalogue.

\begin{figure}
	\centering
	\includegraphics[width=8.4cm,height=5.0cm]{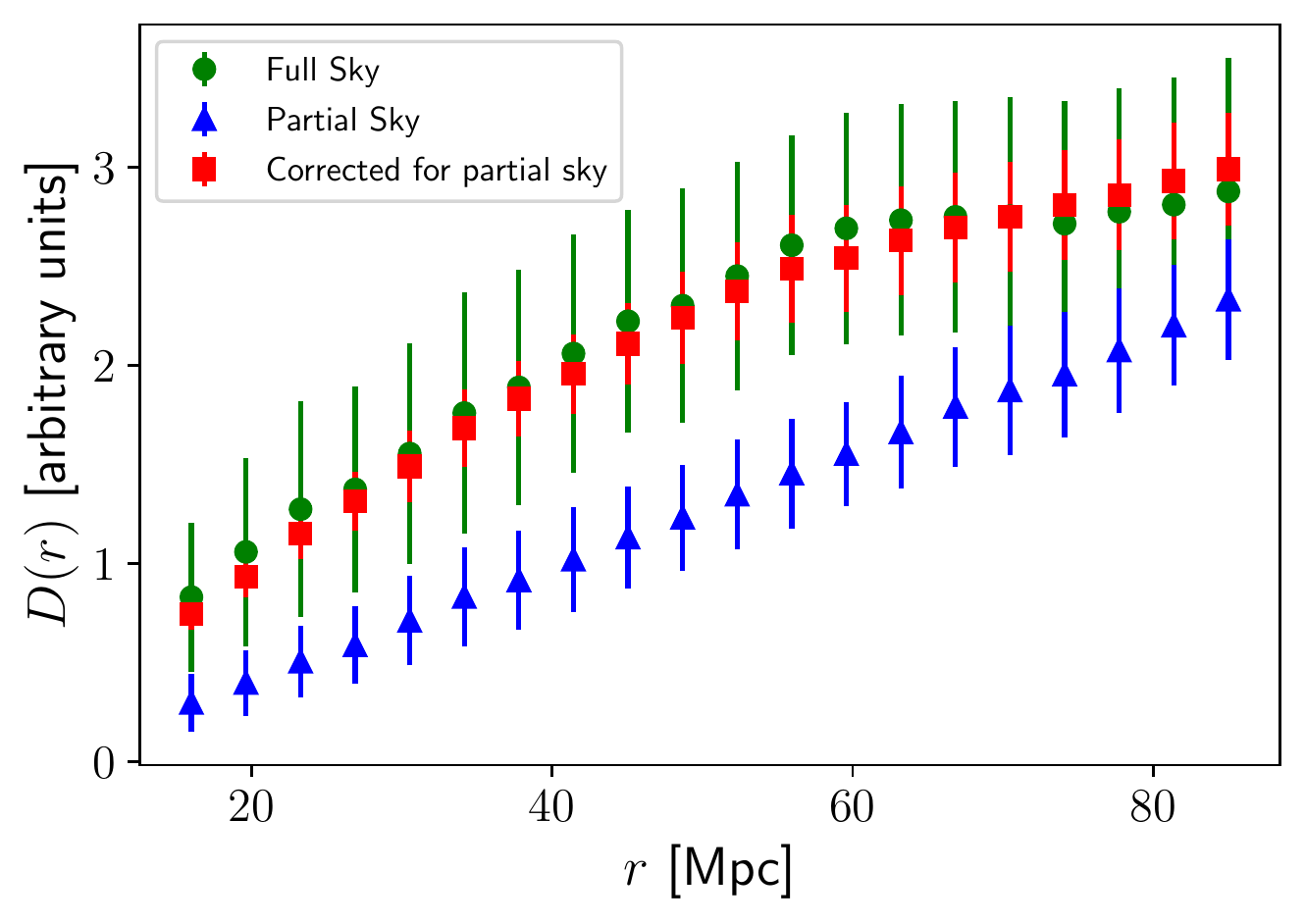}
	\caption{
		Robustness test of the correction procedure, performed for $N=144$ log-normal 
		catalogues. 
		The results show an excellent agreement between the corrected clustering dipoles for the 
		PS catalogues (red squares) as compared with the clustering dipoles of the corresponding 
		FS catalogues (green dots). 
		The average of the uncorrected PS dipoles is represented by blue triangles.
	}
	\label{fig:teste_nulo}
\end{figure}

In the third step we perform the average of these $N$ uncorrected and 
$N$ corrected clustering dipoles and plot them in Fig.~\ref{fig:teste_nulo}.
To complete the test one has to calculate the clustering dipole of the $N$ FS catalogues, 
take their average and plot together with the above data. 
One clearly observes in Fig. \ref{fig:teste_nulo} that our procedure to correct the clustering 
dipole of the PS simulated catalogues perfectly reproduces the true result.

As a complementary verification, we also test our correction procedure by 
calculating the misalignment, $\Delta\theta(r)$, as a function of the radial distance, between 
the LG velocity relative to the CMB frame and the clustering dipole measured from the 
ALFALFA catalogue. 
The expected behaviour for $\Delta\theta(r)$ is a decreasing function for large distances, 
achieving a convergence that depends on the size and location of the data sample on the 
sky, besides of the deepness of the catalogue (see section 5 of \cite{Bilicki11}).

\begin{figure}
	\centering
	\includegraphics[width=8.0cm,height=5.0cm]{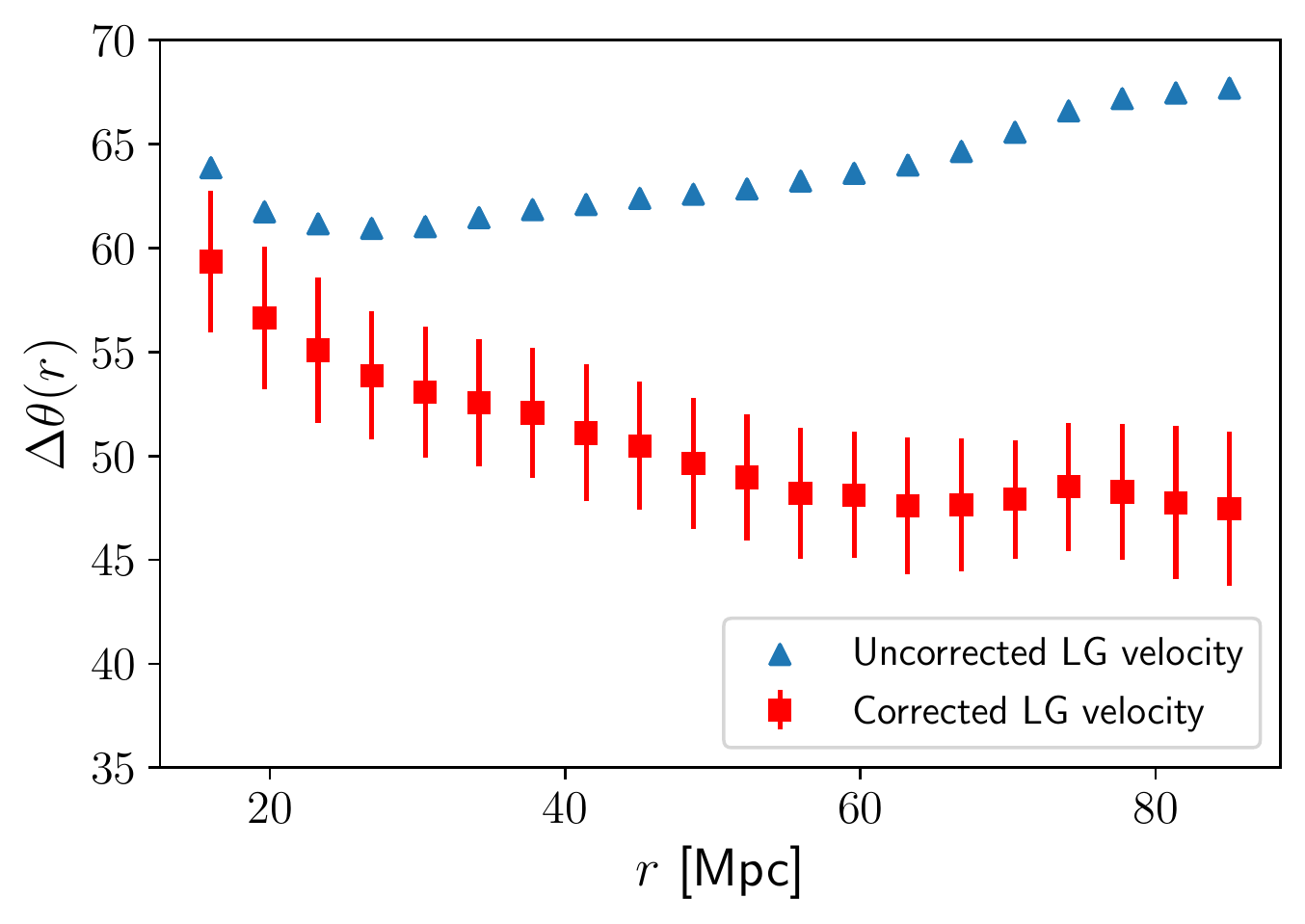}
	\caption{Misalignment between the LG velocity relative to the CMB
		frame, $\textbf{u}_{\text{LG}}$, and the clustering dipole measured from the ALFALFA 
		catalogue, $\textbf{v}_{\text{LG}}(r)\beta^{-1}$,
		for both cases, the corrected (red squares) and uncorrected (blue triangles) velocities.
	}
	\label{fig:desalinhamento}
\end{figure}

We perform the calculation of $\Delta\theta(r)$ with two datasets: 
the original or uncorrected LG velocity and the corrected LG velocity obtained according to 
our correction procedure described above. 
Our results are shown in Fig. \ref{fig:desalinhamento}, where we observe the uncorrected 
(blue triangles) and the corrected (red squares) $\Delta\theta$ as a function of the radial distance, $r$. 
For the uncorrected LG velocity data, the misalignment varies between $60^{\circ}$ and 
$70^{\circ}$, increasing at large scales, very different from a decreasing expected behaviour. 
For the corrected LG velocity data, the misalignment decreases and at large scales 
converges to $\sim 45^{\circ}$.

\section{Testing the Eisenstein \& Hu approach} \label{apendiceD}

\begin{figure}
	\centering
	\includegraphics[width=8.0cm,height=5.0cm]{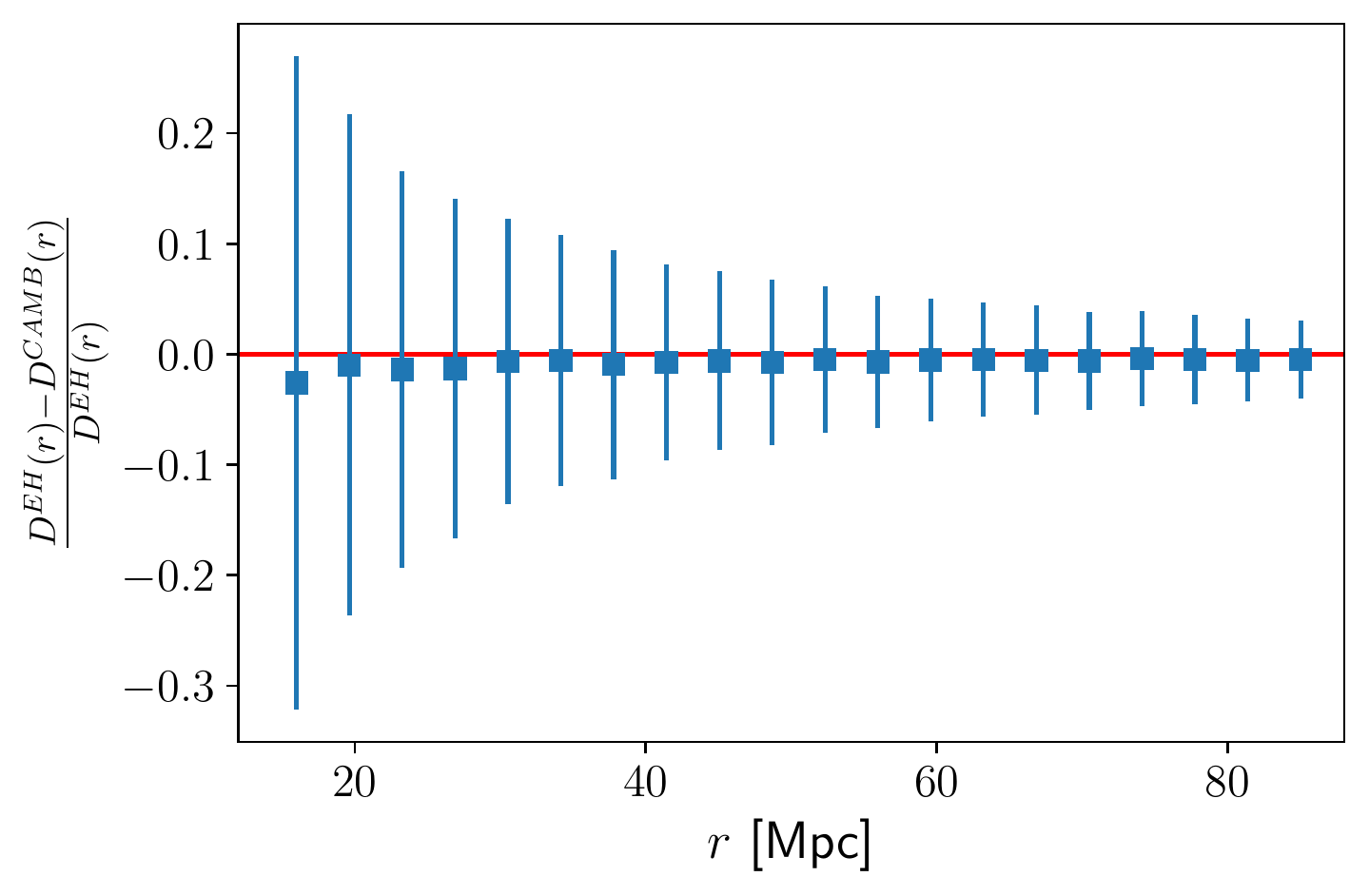}
	\caption{Relative difference of the LG dipole calculated from 1000 log-normal simulations generated using the power spectrum 
		from two approaches: EH~\citep{Eisenstein98} and CAMB~\citep{Lewis00}. 
		The horizontal axis refers to the radial distance from the observer to a surface 
		of radius $r$, where the dipole is calculated.}
	\label{fig:diff_Pk}
\end{figure}

The code of~\citet{Agrawal17}, used here to produce the log-normal simulations, calculates 
the matter power spectrum, $P(k)$, using the Eisenstein \& Hu (EH) transfer 
function~\citep{Eisenstein98}. 
We find interesting to perform a test to check the accuracy of the EH approach compared 
with the result obtained using the CAMB code\footnote{\url{https://camb.info/}}\citep{Lewis00}, 
one of the most known and tested Boltzmann codes. 
For this, first we produce a set of 1000 log-normal simulations using the whole pipeline of 
the code, this includes the internal use of the EH fitting to obtain the matter power spectrum. 
Second, we generate a set of 1000 log-normal simulations, but this time the matter power 
spectrum is produced with the CAMB code and introduced into the code as a numerical 
table (this is an option of the code). 
Then we perform dipole clustering analyses with both sets of log-normal simulations, our 
results are shown in Fig. \ref{fig:diff_Pk} and confirm that both approaches provides the 
same output.

\bsp	
\label{lastpage}

\begin{thebibliography}{149}


\bibitem[\protect\citeauthoryear{Abbott et al.}{2018}]{DES}
Abbott, T. M. C., et al., 2018, \href{https://journals.aps.org/prd/abstract/10.1103/PhysRevD.98.043526}{Phys. Rev. D}, 98, 043526, [\href{https://arxiv.org/abs/1708.01530}{arXiv:1708.01530}]

\bibitem[\protect\citeauthoryear{Adam \& Blake}{2017}]{Adam17}
Adams, C., \& Blake, C., 2017, \href{https://academic.oup.com/mnras/article-abstract/471/1/839/3871371?redirectedFrom=fulltext}{MNRAS}, 471, 839, [\href{https://arxiv.org/abs/1706.05205}{arXiv:1706.05205}]


\bibitem[\protect\citeauthoryear{Agrawal et al.}{2017}]{Agrawal17}
Agrawal, A., et al., 2017, \href{https://iopscience.iop.org/article/10.1088/1475-7516/2017/10/003}{J. Cosmol. Astropart. P.}, 10, 003, [\href{https://arxiv.org/abs/1706.09195}{arXiv:1706.09195}]

\bibitem[\protect\citeauthoryear{Alam et al.}{2017}]{Alam17}
Alam, S., et al., 2017, \href{https://academic.oup.com/mnras/article-abstract/470/3/2617/3091741?redirectedFrom=fulltext}{MNRAS}, 470, 2617, [\href{https://arxiv.org/abs/1607.03155}{arXiv:1607.03155}]

\bibitem[\protect\citeauthoryear{Aubert et al.}{2020}]{Aubert20}
Aubert, M., et al., 2020, [preprint \href{https://arxiv.org/abs/2007.09013}{arXiv:2007.09013}]

\bibitem[\protect\citeauthoryear{Avila et al.}{2018}]{Avila18}
Avila, F., et al., 2018, 
\href{https://iopscience.iop.org/article/10.1088/1475-7516/2018/12/041}
{J. Cosmol. Astropart. P.}, 12, 041, 
[\href{https://arxiv.org/abs/1806.04541}{arXiv:1806.04541}]

\bibitem[\protect\citeauthoryear{Avila et al.}{2019}]{Avila19}
Avila, F., et al., 2019, 
\href{https://academic.oup.com/mnras/article-abstract/488/1/1481/5526236?redirectedFrom=fulltext}{MNRAS}, 488, 1481,
[\href{https://arxiv.org/abs/1906.10744}{arXiv:1906.10744}]



\bibitem[\protect\citeauthoryear{Basilakos \& Plionis}{1998}]{Basilakos98}
Basilakos, S., \& Plionis, M., 1998, \href{https://academic.oup.com/mnras/article/299/3/637/1070528}{MNRAS}, 299, 637, [\href{https://arxiv.org/abs/astro-ph/9801271}{astro-ph/9801271}]

\bibitem[\protect\citeauthoryear{Basilakos \& Plionis}{2006}]{Basilakos06}
Basilakos, S., \& Plionis, M., 2006, \href{https://academic.oup.com/mnras/article/373/3/1112/1062299}{MNRAS}, 373, 1112, [\href{https://arxiv.org/abs/astro-ph/0609476}{astro-ph/0609476}]

\bibitem[\protect\citeauthoryear{Basilakos}{2012}]{Basilakos12}
Basilakos, S., 2012, \href{https://www.worldscientific.com/doi/abs/10.1142/S0218271812500642}{Int. J. Mod. Phys. D}, 21, 1250064, [\href{https://arxiv.org/abs/1202.1637}{arXiv:1202.1637}]

\bibitem[\protect\citeauthoryear{Bautista et al.}{2018}]{BAODR14}
Bautista, J. E., et al., 2018, \href{https://iopscience.iop.org/article/10.3847/1538-4357/aacea5}{ApJ}, 863, 110, [\href{https://arxiv.org/abs/1712.08064}{arXiv:1712.08064}]

\bibitem[\protect\citeauthoryear{Bilicki et al.}{2011}]{Bilicki11}
Bilicki, M., et al., 2011 , \href{https://iopscience.iop.org/article/10.1088/0004-637X/741/1/31}{ApJ}, 741, 31, [\href{https://arxiv.org/abs/1102.4356}{arXiv:1102.4356}]

\bibitem[\protect\citeauthoryear{Bilicki}{2012}]{Bilickithesis}
Bilicki, M., 2012, [preprint \href{https://arxiv.org/abs/1205.1970}{1205.1970v1}]



\bibitem[\protect\citeauthoryear{Boruah, Hudson \& Lavaux}{2019}]{Boruah}
Boruah, S. S., Hudson, M. J., \& Lavaux, G., 2019, [preprint \href{https://arxiv.org/abs/1912.09383}{arXiv:1912.09383}]



\bibitem[\protect\citeauthoryear{Courteau \& Van Den Bergh}{1999}]{Courteau}
Courteau, S., \& Van Den Bergh, S, 1999, \href{https://iopscience.iop.org/article/10.1086/300942}{AJ}, 118, 337, [\href{https://arxiv.org/abs/astro-ph/9903298}{astro-ph/9903298}]




\bibitem[\protect\citeauthoryear{de Carvalho et al.}{2018}]{deCarvalho18}
de Carvalho, E., et al.,  2018, \href{https://iopscience.iop.org/article/10.1088/1475-7516/2018/04/064}{J. Cosmol. Astropart. P.}, 04, 064, [\href{https://arxiv.org/abs/1709.00113}{arXiv:1709.00113}]

\bibitem[\protect\citeauthoryear{de Carvalho et al.}{2020}]{deCarvalho20}
de Carvalho, E., et al., 2020, 
\href{https://academic.oup.com/mnras/article-abstract/492/3/4469/5707432?redirectedFrom=fulltext}
{MNRAS}, 492, 4469,
[\href{https://arxiv.org/abs/2002.01109}{arXiv:2002.01109}]

\bibitem[\protect\citeauthoryear{de Carvalho et al.}{2021}]{deCarvalho21}
de Carvalho, E., et al., 2021, 
\href{https://www.aanda.org/articles/aa/abs/2021/05/aa39936-20/aa39936-20.html}
{A\&A}, 649, A20,
[\href{https://arxiv.org/pdf/2103.14121}{arXiv:2103.14121}]


\bibitem[\protect\citeauthoryear{Eisenstein \& Hu}{1998}]{Eisenstein98}
Eisenstein, D. J., \& Hu, W., 1998, \href{https://iopscience.iop.org/article/10.1086/305424}{ApJ}, 496, 605, [\href{https://arxiv.org/abs/astro-ph/9709112}{arXiv:astro-ph/9709112}]


\bibitem[\protect\citeauthoryear{Erdo\v{g}du et al.}{2006}]{Erdogdu}
Erdo\v{g}du, P., et al., 2006, \href{https://academic.oup.com/mnras/article/368/4/1515/1151639}{MNRAS}, 368, 1515, [\href{https://arxiv.org/abs/astro-ph/0507166}{astro-ph/0507166}]





\bibitem[\protect\citeauthoryear{Giovanelli \& Haynes}{2016}]{Giovanelli}
Giovanelli, R., \& Haynes, M. P., 2016, \href{https://link.springer.com/article/10.1007/s00159-015-0085-3}{Astron Astrophys Rev}, 24, 1, [\href{https://arxiv.org/abs/1510.04660}{arXiv:1510.04660}]





\bibitem[\protect\citeauthoryear{Haynes et al.}{2011}]{Haynes11}
Haynes, M. P., et al., 2011, \href{https://iopscience.iop.org/article/10.1088/0004-6256/142/5/170}{AJ}, 142, 170, [\href{https://arxiv.org/abs/1109.0027}{arXiv:1109.0027}]

\bibitem[\protect\citeauthoryear{Haynes et al.}{2018}]{Haynes18}
Haynes, M. P., et al., 2018, \href{https://iopscience.iop.org/article/10.3847/1538-4357/aac956}{ApJ}, 861, 49, [\href{https://arxiv.org/abs/1805.11499}{arXiv:1805.11499}]

\bibitem[\protect\citeauthoryear{Haude et al.}{2019}]{Haude19}
Haude, S., et al., 2019, [preprint \href{https://arxiv.org/abs/1912.04560}{1912.04560}]

\bibitem[\protect\citeauthoryear{Heinesen}{2020}]{Heinesen20}
Heinesen, A., 2020, \href{https://iopscience.iop.org/article/10.1088/1475-7516/2020/10/052}{J. Cosmol. Astropart. P.}, 10, 052, [\href{https://arxiv.org/abs/2006.15022}{arXiv:2006.15022}]

\bibitem[\protect\citeauthoryear{Howlett et al.}{2017}]{Howlett17}
Howlett, C., et al., 2017, \href{https://academic.oup.com/mnras/article-abstract/471/3/3135/3873951?redirectedFrom=fulltext}{MNRAS}, 471, 3135, [\href{https://arxiv.org/abs/1706.05130}{arXiv:1706.05130}]

\bibitem[\protect\citeauthoryear{Hudson}{1993}]{Hudson93}
Hudson, M. J., 1993, \href{https://academic.oup.com/mnras/article/265/1/72/975255}{MNRAS}, 265, 72

\bibitem[\protect\citeauthoryear{Huterer et al.}{2017}]{Huterer17}
Huterer, D., et al., 2017, \href{https://iopscience.iop.org/article/10.1088/1475-7516/2017/05/015}{J. Cosmol. Astropart. P.}, 5, 015, [\href{https://arxiv.org/abs/1611.09862}{arXiv:1611.09862}]



\bibitem[\protect\citeauthoryear{Jones et al.}{2016}]{Jones16}
Jones, M. G., et al., 2016, \href{https://academic.oup.com/mnras/article/457/4/4393/2589071}{MNRAS}, 457, 4393, [\href{https://arxiv.org/abs/1510.07050}{arXiv:1510.07050}]

\bibitem[\protect\citeauthoryear{Jones et al.}{2018}]{Jones18}
Jones, M. G., et al., 2018, \href{https://academic.oup.com/mnras/article-abstract/477/1/2/4911535?redirectedFrom=fulltext}{MNRAS}, 477, 2, [\href{https://arxiv.org/abs/1802.00053}{arXiv:1802.00053}]

\bibitem[\protect\citeauthoryear{Jones et al.}{2020}]{Jones20}
Jones, M. G., et al., 2020, \href{https://academic.oup.com/mnras/article-abstract/494/2/2090/5815101?redirectedFrom=fulltext}{MNRAS}, 494, 2090, [\href{https://arxiv.org/abs/2003.09302}{arXiv:2003.09302}]


\bibitem[\protect\citeauthoryear{Juszkiewicz et al.}{2009}]{Juszkiewicz}
Juszkiewicz, R., et al., 2009, \href{https://iopscience.iop.org/article/10.1088/1475-7516/2010/02/021}{J. Cosmol. Astropart. P.}, 2, 021, [\href{https://arxiv.org/abs/0901.0697}{arXiv:0901.0697v3}]




\bibitem[\protect\citeauthoryear{Kaiser}{1987}]{Kaiser87}
Kaiser, N., 1987, \href{https://academic.oup.com/mnras/article/227/1/1/1065830}{MNRAS}, 227, 1

\bibitem[\protect\citeauthoryear{Kaiser \& Lahav}{1989}]{Kaiser89}
Kaiser, N., \& Lahav, O., 1989, \href{https://academic.oup.com/mnras/article/237/1/129/1068507}{MNRAS}, 237, 129

\bibitem[\protect\citeauthoryear{Kocevski \& Ebeling}{2006}]{Kocevski}
Kocevski, D. D., \& Ebeling, H., 2006, \href{https://iopscience.iop.org/article/10.1086/503666}{ApJ}, 645, 1043, [\href{https://arxiv.org/abs/astro-ph/0510106}{astro-ph/0510106}]

\bibitem[\protect\citeauthoryear{Kogut et al.}{1993}]{Kogut}
Kogut, A., et al., 1993, \href{http://articles.adsabs.harvard.edu/pdf/1993ApJ...419....1K}{ApJ}, 419, 1, [\href{https://arxiv.org/abs/astro-ph/9312056}{astro-ph/9312056}]

\bibitem[\protect\citeauthoryear{Kolokotronis et al.}{1995}]{Kolokotronis}
Kolokotronis, V., et al., 1996, \href{http://articles.adsabs.harvard.edu/pdf/1996MNRAS.280..186K}{MNRAS}, 280, 186, [\href{https://arxiv.org/abs/astro-ph/9511053}{arXiv:astro-ph/9511053}]



\bibitem[\protect\citeauthoryear{Lewis, Challinor \& Cahn}{2000}]{Lewis00}
Lewis, A., Challinor, A., \& Lasenby, A., 2000, \href{https://iopscience.iop.org/article/10.1086/309179}{ApJ}, 538, 473, [\href{https://arxiv.org/abs/astro-ph/9911177}{arXiv:astro-ph/9911177}]


\bibitem[\protect\citeauthoryear{Linder \& Cahn}{2007}]{Linder}
Linder, E. V., \& Cahn, R. N., 2007, \href{https://www.sciencedirect.com/science/article/abs/pii/S0927650507001326?via\%3Dihub}{Astropart. Phys.}, 28, 481, [\href{https://arxiv.org/abs/astro-ph/0701317}{astro-ph/0701317v2}]





\bibitem[\protect\citeauthoryear{Marques et al.}{2018}]{Marques18}
Marques, G. A., Novaes, C. P., Bernui A., Ferreira, I. S., 2018, 
\href{https://academic.oup.com/mnras/article-abstract/473/1/165/4103556?redirectedFrom=fulltext}
{MNRAS}, 473, 165, 
[\href{https://arxiv.org/abs/1708.09793}{arXiv:1708.09793}]

\bibitem[\protect\citeauthoryear{Marques \& Bernui}{2020}]{Marques20}
Marques, G. A., \& Bernui, A., 2020, \href{https://iopscience.iop.org/article/10.1088/1475-7516/2020/05/052}{J. Cosmol. Astropart. P.}, 5, 052, [\href{https://arxiv.org/abs/1908.04854}{arXiv:1908.04854}]

\bibitem[\protect\citeauthoryear{Martin et al.}{2012}]{Martin12}
Martin, A. M., et al., 2012, \href{https://iopscience.iop.org/article/10.1088/0004-637X/750/1/38}{ApJ}, 750, 38, [\href{https://arxiv.org/abs/1202.6005}{arXiv:1202.6005v1}]


\bibitem[\protect\citeauthoryear{Master}{2005}]{Master}
Master, K, L., 2005, PhD thesis, Univ. Cornell



\bibitem[\protect\citeauthoryear{Nunes \& Bernui}{2020}]{Nunes20}
Nunes, R. C., \& Bernui, A., 2020, \href{https://link.springer.com/article/10.1140\%2Fepjc\%2Fs10052-020-08601-8}{Eur. Phys. J. C}, 80, 1025, [\href{https://arxiv.org/abs/2008.03259}{arXiv:2008.03259}]



\bibitem[\protect\citeauthoryear{O'Donoghue}{2018}]{Donoghue}
O'Donoghue, A. A., et al., 2018, [preprint \href{https://arxiv.org/abs/1811.01283}{arXiv:1811.01283}]




\bibitem[\protect\citeauthoryear{Pandey \& Sarkar}{2020}]{Pandey20}
Pandey, B., \& Sarkar, S., 2020, \href{https://academic.oup.com/mnras/article-abstract/498/4/6069/5909614?redirectedFrom=fulltext}{MNRAS}, 498, 6069, [\href{https://arxiv.org/abs/2002.08400}{arXiv:2002.08400}]

\bibitem[\protect\citeauthoryear{Papageorgiou et al.}{2012}]{Papageorgiou}
Papageorgiou, A., et al., \href{https://academic.oup.com/mnras/article/422/1/106/1019897}{MNRAS}, 2012, 422, 106, [\href{https://arxiv.org/abs/1201.4878}{arXiv:1201.4878}]

\bibitem[\protect\citeauthoryear{Papastergis et al.}{2013}]{Papastergis}
Papastergis, E., et al., 2013, \href{https://iopscience.iop.org/article/10.1088/0004-637X/776/1/43}{ApJ}, 776, 43, [\href{https://arxiv.org/abs/1308.2661}{arXiv:1308.2661v1}]

\bibitem[\protect\citeauthoryear{Peebles}{1980}]{peebles}
Peebles, P. J. E., 1980, The large-scale structure of the universe. Princeton Univ. Press, Princeton, NJ

\bibitem[\protect\citeauthoryear{Pezzotta et al.}{2017}]{Pezotta}
Pezzotta, A., et al., 2017, \href{https://www.aanda.org/articles/aa/abs/2017/08/aa30295-16/aa30295-16.html}{A\&A}, 604, A33, [\href{https://arxiv.org/abs/1612.05645}{arXiv:1612.05645}]

\bibitem[\protect\citeauthoryear{Planck Collaboration}{2016}]{Planck16}
Planck Collaboration, 2016, 
\href{https://www.aanda.org/articles/aa/abs/2016/10/aa25833-15/aa25833-15.html}{A\&A}, 594, A24, [\href{https://arxiv.org/abs/1502.01597}{arXiv:1502.01597}]

\bibitem[\protect\citeauthoryear{Planck Collaboration}{2020}]{Planck18}
Planck Collaboration, 2020, 
\href{https://www.aanda.org/articles/aa/abs/2020/09/aa33910-18/aa33910-18.html}{A\&A}, 641, 
A6, [preprint, \href{https://arxiv.org/abs/1807.06209}{arXiv:1807.06209}]




\bibitem[\protect\citeauthoryear{Qin, Howlett, \& Staveley-Smith}{2019}]{Qin19}
Qin, F., Howlett, C., \& Staveley-Smith, L., 2019, \href{https://academic.oup.com/mnras/article-abstract/487/4/5235/5513479?redirectedFrom=fulltext}{MNRAS}, 487, 5235, [\href{https://arxiv.org/abs/1906.02874}{arXiv:1906.02874}]



\bibitem[\protect\citeauthoryear{Riess et al.}{2019}]{riess2019}
Riess, A. G., et al., 2019, \href{https://iopscience.iop.org/article/10.3847/1538-4357/ab1422}{ApJ}, 876, 85, [\href{https://arxiv.org/abs/1903.07603}{arXiv:1903.07603}]

\bibitem[\protect\citeauthoryear{Rowan-Robinson et al.}{2000}]{rowan}
Rowan-Robinson, M., et al., 2000, \href{https://academic.oup.com/mnras/article/314/2/375/1173924}{MNRAS}, 314, 375, [\href{https://arxiv.org/abs/astro-ph/9912223}{arXiv:astro-ph/9912223}]


\bibitem[\protect\citeauthoryear{Said et al.}{2020}]{Said20}
Said, K., et al., 2020, \href{https://academic.oup.com/mnras/article-abstract/497/1/1275/5870121?redirectedFrom=fulltext}{MNRAS}, 497, 1275, [\href{https://arxiv.org/abs/2007.04993}{arXiv:2007.04993}]

\bibitem[\protect\citeauthoryear{Sarkar \& Pandey}{2019}]{Sarkar19}
Sarkar, S., \& Pandey, B., 2019, \href{https://academic.oup.com/mnras/article-abstract/485/4/4743/5421581?redirectedFrom=fulltext}{MNRAS}, 485, 4743, [\href{https://arxiv.org/abs/1812.03661}{arXiv:1812.03661}]

\bibitem[\protect\citeauthoryear{Scaramella, Vettolani \& Zamorani}{1994}]{Scaramella}
Scaramella, R., Vettolani, G., \& Zamorani, G., 1994, \href{http://articles.adsabs.harvard.edu/pdf/1994ApJ...422....1S}{ApJ}, 422, 1

\bibitem[\protect\citeauthoryear{Schmoldt et al.}{1999}]{Schmoldt}
Schmoldt, I., et al., 1999, \href{https://academic.oup.com/mnras/article/304/4/893/1048735}{MNRAS}, 304, 893, [\href{https://arxiv.org/abs/astro-ph/9901087}{astro-ph/9901087}]

\bibitem[\protect\citeauthoryear{Springob et al.}{2007}]{Spring07}
Springob, C. M., et al., 2007, \href{https://iopscience.iop.org/article/10.1088/0067-0049/182/1/474}{ApJS}, 172, 599, [\href{https://arxiv.org/abs/0705.0647}{arXiv:0705.0647}]

\bibitem[\protect\citeauthoryear{Strauss \& Willick}{1995}]{Strauss}
Strauss, M. A., \& Willick, J. A., 1995, \href{https://www.sciencedirect.com/science/article/abs/pii/0370157395000137?via\%3Dihub}{Phys. Rep.}, 261, 271, [\href{https://arxiv.org/abs/astro-ph/9502079}{astro-ph/9502079}]



\bibitem[\protect\citeauthoryear{Turnbull et al.}{2012}]{Turnbull}
Turnbull, S. J., 2012, \href{https://academic.oup.com/mnras/article/420/1/447/1046487}{MNRAS}, 420, 447, [\href{https://arxiv.org/abs/1111.0631}{arXiv:1111.0631}]



\bibitem[\protect\citeauthoryear{Van den Bergh}{2000}]{Berghbook}
Van den Bergh, S., 2000, The galaxies of the Local Group. Cambridge Univ. Press, Cambridge, UK

\bibitem[\protect\citeauthoryear{Van der Marel \& Guhathakurta}{2008}]{Marel}
Van der Marel, R. P., \& Guhathakurta, P., 2008, \href{https://iopscience.iop.org/article/10.1086/533430}{ApJ}, 678, 187, [\href{https://arxiv.org/abs/0709.3747}{arXiv:0709.3747}]







\end{thebibliography}
\end{document}